\documentclass[aps,prfluids,preprint,amsmath,amssymb,superscriptaddress,nofootinbib]{revtex4-2}
\usepackage{graphicx}
\usepackage{xcolor}
\usepackage{tikz}
\usepackage{subcaption}
\usepackage{placeins}
\usetikzlibrary{arrows.meta, calc, quotes, angles,patterns, patterns.meta}
\definecolor{markstrong}{HTML}{111111}
\definecolor{markasym}{HTML}{444444}
\definecolor{markbypass}{HTML}{999999}
\definecolor{marksecondmode}{HTML}{F4B6B6}

\DeclareRobustCommand{\legstrong}{\raisebox{-0.2ex}{\tikz{\fill[markstrong] (0,0) circle (0.09);}}}
\DeclareRobustCommand{\legasym}{\raisebox{-0.2ex}{\tikz{\draw[markasym, line width=0.5pt] (-0.09,-0.09) rectangle (0.09,0.09);}}}
\DeclareRobustCommand{\legbypass}{\raisebox{-0.2ex}{\tikz{\draw[markbypass, line width=0.5pt, dash pattern=on 1pt off 1pt] (0,0) circle (0.09);}}}
\DeclareRobustCommand{\legsecondmode}{\raisebox{-0.2ex}{\tikz{\fill[marksecondmode] (0,0) circle (0.13); \fill[markstrong] (0,0) circle (0.09);}}}

\begin{document}
\title{Hydrodynamic regimes of the edge-tone in a simplified recorder mouthpiece}

\author{A. Tassawar}
\affiliation{Dpto.~Ing. Mec\'anica y Dise\~no Industrial, Escuela Superior de Ingenier\'ia, Universidad de C\'adiz, Av.~de la Universidad de C\'adiz 10, 11519 Puerto Real, Espa\~na}
\author{R. Velasco-Segura}
\affiliation{Instituto de Ciencias Aplicadas y Tecnolog\'ia, Universidad Nacional Aut\'onoma de M\'exico, Circuito Exterior S/N, Ciudad Universitaria, 04510, Mexico City}
\author{P. Rendon}
\affiliation{Instituto de Ciencias Aplicadas y Tecnolog\'ia, Universidad Nacional Aut\'onoma de M\'exico, Circuito Exterior S/N, Ciudad Universitaria, 04510, Mexico City}
\author{M. Fosas de Pando}
\affiliation{Dpto.~Ing. Mec\'anica y Dise\~no Industrial, Escuela Superior de Ingenier\'ia, Universidad de C\'adiz, Av.~de la Universidad de C\'adiz 10, 11519 Puerto Real, Espa\~na}
\author{R. Ostilla-M\'onico}
\email[Contact author: ]{rodolfo.ostilla@uca.es}
\affiliation{Dpto.~Ing. Mec\'anica y Dise\~no Industrial, Escuela Superior de Ingenier\'ia, Universidad de C\'adiz, Av.~de la Universidad de C\'adiz 10, 11519 Puerto Real, Espa\~na}

\date{\today}

\begin{abstract}
The acoustic mechanism of flue instruments originates in an aerodynamic process in which an air jet interacts with a sharp edge, the labium. Previous studies have focused on isolated edge-tones or on full instruments in which acoustic resonance obscures the underlying hydrodynamics, and a systematic understanding of how geometric control parameters govern jet stability remains lacking. This study presents a large-scale parametric study of the jet--labium interaction in a simplified recorder mouthpiece without a resonator, using two-dimensional numerical simulations. The incompressible Navier--Stokes equations are solved by direct numerical simulation with an immersed boundary method, in more than 200 simulations that vary four parameters: the mouth distance, the labium angle, the vertical offset of the labium, and the Reynolds number ($Re$). The dynamics are characterized through the spectra of the transverse force on the labium. The baseline configuration oscillates as a predominantly periodic Stage I edge-tone  with a Strouhal number $St \approx 0.07$. This value remains nearly constant for $Re \ge 330$ and for all labium angles at the baseline mouth distance, so the frequency is set primarily by the geometry and not by viscosity. The mouth distance has the strongest influence on the flow, and it is the only parameter that introduces a second oscillation stage. The frequency of the Stage I peak is inversely proportional to the mouth distance. The second stage is weakly present at five and six windway heights, clearly established from six and a half, and dominant from eight, where it produces strong amplitude modulation; at the largest distance studied the force is highly irregular. At the baseline mouth distance, the vertical offset acts mainly on the strength of the oscillation: small offsets leave it unchanged, moderate offsets make it more variable, and an offset of half a windway height strongly reduces its amplitude. The labium angle does not change the underlying instability mechanism. A baseline configuration is identified for future aeroacoustic coupling studies.
\end{abstract}

\maketitle

\section{Introduction}
The acoustic mechanism of flue instruments has long remained a challenging
problem in musical acoustics~\cite{fletcher1998}, and despite decades of investigation, many
aspects of the underlying flow physics are still not fully understood. The
sound in these instruments originates from an aerodynamic process in which a
thin air jet emerging from a narrow windway interacts with a sharp edge known
as the labium~\cite{coltman1968}. The oscillatory deflection of this jet produces pressure
fluctuations that couple with the resonant air
column to sustain tonal sound production. Because the stability of the
oscillation and the characteristics of the emitted sound depend strongly on
the jet behavior near the mouth, understanding the fluid dynamics of this
region in isolation from the downstream resonator is essential. This is the
aim of the present study.

The fundamental aerodynamic configuration underlying sound generation in flue
instruments is the \emph{edge-tone }: a planar free jet impinging on a
wedge-shaped edge that, under certain conditions, develops self-sustained
oscillations without the presence of any resonating cavity. It belongs to the
wider class of self-sustained oscillations of impinging free shear
layers~\cite{rockwell1979}. This
configuration has been studied for nearly two centuries, since Sondhauss first
noted in 1854~\cite{sondhauss1854ueber} that audible tones could be produced by blowing a jet against
an edge without a resonating column. The edge-tone  represents the
canonical aeroacoustic building block of flue instrument acoustics, and a
thorough understanding of its behavior is a prerequisite for understanding
the more complex dynamics of complete instruments.

The first systematic experimental investigations of the edge-tone 
established its most characteristic features. Brown~\cite{brown1937} conducted careful
quantitative experiments on a jet issuing from a narrow slit and impinging on a
wedge, documenting four distinct oscillation stages\footnote{
The phenomenon described by Brown and many other subsequent studies as
``stages'' is in some recent publications referred to as ``modes,'' or ``hydrodynamic modes.'' In this work we choose to keep the term ``stages.''}, the hysteresis between
them, and the coexistence of multiple frequencies. He found
that the oscillation frequency varies systematically with jet velocity and
edge-to-nozzle distance, and proposed a semi-empirical formula capturing
these dependencies. It was also noted
that the whole edge-tone phenomenon occurs
within the sound-sensitive range of the jet, and that at low speeds the
feedback is predominantly hydrodynamic while acoustic effects become
significant at higher speeds~\cite{brown1937b}. Powell~\cite{powell1961} later related sound production to the
unsteady motion of vortices within the jet, emphasizing vortex shedding as a
primary driver of self-sustained oscillations, attributed the sound to the
dipole associated with the fluctuating force on the edge, formulated a gain
criterion for the oscillation, and developed a feedback-loop model in which
a dipole sound source at the edge re-excites the jet at the
nozzle exit.

Subsequent theoretical and experimental work progressively refined this
picture. Curle~\cite{curle1953} developed a purely hydrodynamic vortex theory in which
alternating vortices are produced at the nozzle and at the edge
tip, while Holger \textit{et al.}~\cite{holger1977} proposed a vortex-street model
that treated the edge as a flat plate and derived a phase criterion for the
oscillation. Nyborg~\cite{nyborg1954} presented a dynamic theory that
described the shape of the oscillating jet centerline and predicted frequency
ratios between stages. Crighton~\cite{crighton1992} later derived a linear theory for
the frequencies of the operating stages of the jet edge-tone feedback cycle,
and Kaykayoglu and Rockwell~\cite{kaykayoglu1986vortex} measured the
instantaneous pressure fields during unstable jet--edge interaction.
A recurring point of debate across
these studies was the exponent governing the dependence of oscillation
frequency on nozzle-to-edge distance, with early theoretical treatments
predicting an exponent of $3/2$ while experimental evidence consistently
pointed toward a value closer to $1$.

More recently, Vaik, Varga and Pa\'{a}l~\cite{vaik2014a,vaik2014b} carried out a detailed parametric study
of the edge-tone  by both computational and experimental means, covering a
wide range of Reynolds numbers and dimensionless nozzle-to-edge distances for
both top-hat and parabolic exit velocity profiles.
Their work confirmed the exponent of $1$ in the frequency scaling law,
clarified the Reynolds number dependence of each oscillation stage, and
showed that the convection velocity of the jet disturbance is not constant
between the nozzle and the edge, contradicting a common assumption in
theoretical models~\cite{holger1977}. They also demonstrated that two-dimensional simulations
of the incompressible Navier--Stokes equations reliably reproduce
experimental results, validating the use of this framework for systematic
parametric studies. These findings built upon earlier incompressible
simulations by Pa\'{a}l and Vaik~\cite{paal2007}, who reproduced three distinct oscillation
stages numerically for the first time and documented the dependence of stage
selection on Reynolds number and nozzle-to-edge
distance. The rich nonlinear dynamics associated with stage
transitions, including spontaneous jumps between stages at fixed flow
conditions and the simultaneous coexistence of multiple stages, were further
characterized by Vaik and Pa\'{a}l~\cite{vaik2011} through combined experimental and numerical
investigations. Bamberger \textit{et al.}~\cite{bamberger2004} similarly combined experiment and
two-dimensional numerical simulation for a parabolic edge-tone configuration,
verifying the linear frequency-distance scaling law and achieving agreement
between experimental and computational Strouhal numbers within
13--15\%. Nagy, Szab\'{o} and Pa\'{a}l~\cite{nagy2021} later proposed a feedback
model for the edge-tone  using linear adjoint stability theory, calculating
the open-loop gain of the oscillation without empirical constants and showing
that a purely hydrodynamic potential vortex at the edge tip is a physically
consistent and energetically sufficient feedback agent at low Mach
numbers.

The edge-tone  thus exhibits rich dynamics, including multiple oscillation
stages, frequency jumps, hysteresis, and coexistence of stages, that are
still not fully understood from first principles. When a
resonating air column is added, as in a recorder or organ pipe, the acoustic
field of the resonator takes over part of the feedback loop and strongly
influences which oscillation regime is selected and at what amplitude.
Semi-empirical theories refined by Coltman~\cite{coltman1968,coltman1976} described the jet-drive mechanism
and addressed how the interaction between the jet and the labium regulates
oscillation regimes in flute-like instruments with a
resonator. These models successfully capture
several observed trends but rely on simplifying assumptions and cannot fully
resolve the nonlinear flow structures that develop near the mouth.
Experimental evidence further highlights the critical role of mouth geometry
in governing the aeroacoustic response of flue systems. Dequand \textit{et al.}~\cite{dequand2003}
conducted a systematic investigation using a simplified whistle configuration
to isolate geometric effects on the sound source. Their
study demonstrated that the labium angle $\alpha$ and the ratio between the mouth
width $W$ and the jet thickness strongly influence both spectral
characteristics and oscillation stability. Recorders, typically characterized
by a sharp labium with edge angles near $15^{\circ}$ and a nearly fixed ratio
of mouth width to jet thickness of about $4$, tend to produce harmonic-rich acoustic fields but are
more sensitive to flow perturbations. In contrast, transverse flutes employ
larger edge angles on the order of $60^{\circ}$, promoting greater
oscillation stability while potentially reducing broadband turbulence noise.
S\'egoufin \textit{et al.}~\cite{segoufin2000} focused on the manner in which the length and geometry of the recorder windway affect the sensitivity of the jet in relation to the acoustic field and, in consequence, influence both the sound produced and the playability of the instrument.
Computational studies of complete flue instruments with resonators have
further expanded understanding of vortex formation and acoustic oscillation
in recorder-like configurations. Early work in this direction was carried
out by Skordos~\cite{skordos1995}. K\"{u}hnelt~\cite{kuhnelt2003} subsequently applied
lattice-Boltzmann simulations to the same class of problems.
Giordano~\cite{giordano2013} first studied a two-dimensional recorder model using the
Navier--Stokes equations, capturing vortex shedding dynamics and the
dependence of the sound spectrum on blowing speed. He
later extended this work to a fully three-dimensional model~\cite{giordano2014}, investigating
the effects of channel geometry, chamfers, and labium position on the attack
transient and steady-state tonal properties. Miyamoto
\textit{et al.}~\cite{miyamoto2013} applied large eddy simulations to further probe these
recorder-like configurations. In contrast, studies of
the isolated edge-tone  without a resonator have continued to refine the
understanding of the fundamental aeroacoustic mechanism. More recently,
Iwagami \textit{et al.}~\cite{iwagami2021} conducted highly resolved compressible direct
numerical simulations of the edge-tone  using an immersed boundary approach,
reproducing the first and second oscillation stages, in agreement with
Brown's equation. These simulations confirmed that the
sound radiation follows a dipole-like fifth-power scaling law for the
two-dimensional configuration.

Despite these advances, a predictive understanding of how geometric control
parameters govern jet stability and sound production at the mouth in the
absence of the resonator remains lacking. Most previous investigations have
either concentrated on the edge-tone  in isolation with simple symmetric wedge
geometries, or on complete instrument configurations where the acoustic
resonance dominates and the underlying hydrodynamics near the mouth are
difficult to isolate. A systematic large-scale parametric study of the
jet--labium interaction in a realistic recorder mouthpiece geometry, without
the resonator, and spanning a broad range of geometric and flow parameters
simultaneously, has not previously been undertaken.

The present work addresses this gap through a parametric study of the
incompressible hydrodynamics of the jet--labium interaction in a simplified
recorder mouthpiece without a resonator. We focus on the region near the mouth,
where the jet issuing from the windway impinges on a single-sided inclined
labium. The two-dimensional incompressible Navier--Stokes equations are solved
using an immersed boundary method, and the mouth distance ($W$), the labium
angle ($\alpha$), the vertical offset ($s$) and the Reynolds number ($Re$) are
varied. The dynamics of each case are characterized through the spectrum and
the time history of the transverse force on the labium. Our aim is to identify
the effect of these parameters on the selection of the oscillation stage and on
its amplitude. To the authors' knowledge, this is the first systematic study of
this interaction in a recorder-type mouthpiece that varies these four
parameters, in sweeps about a baseline configuration.

The paper is organized as follows. Section~\ref{sec:methodology} describes the
model geometry and boundary conditions, the numerical method, the ranges of the
control parameters, the spatial and temporal resolution, and the spectral
analysis. Section~\ref{sec:results} presents the results: it first describes the
flow regimes at the lowest Reynolds numbers and justifies the use of the
transverse labium force as the diagnostic quantity, then examines in turn the
effects of the Reynolds number, the mouth distance, the vertical offset and the
labium angle, and ends with regime maps in the $(s,\alpha)$ and $(s,W)$ planes.
Section~\ref{sec:conclusion} summarizes the findings and proposes a baseline
configuration for future aeroacoustic studies. Appendix~\ref{app:convergence}
verifies the numerical parameters.

\section{Methodology}
\label{sec:methodology}

\subsection{Model Geometry}
\label{subsec:geometry}

\begin{figure}[!htbp]
    \centering
    \begin{subfigure}[b]{0.42\textwidth}
        \centering
        \resizebox{\textwidth}{!}{%
        \begin{tikzpicture}[
            >={Latex[length=1.5mm, width=0.8mm]},
            dim/.style={black, thin, <->},
            dim_label/.style={text=black, font=\sffamily\small},
            ext/.style={black, thin}
        ]
            \draw[black, dotted, thick] (0,0) rectangle (12,8);
            \node[below=6pt, font=\sffamily\large\itshape] at (6,0) {$x$};
            \node[left=6pt, font=\sffamily\large\itshape] at (0,4) {$y$};

            \begin{scope}
                \clip (9.5,0) rectangle (12,8);
                \foreach \i in {-8,-7.3,...,12} {
                    \draw[black, thin, dashed] (\i, -1) -- (\i+8, 9);
                }
            \end{scope}
            \draw[black, dotted, thick] (9.5,0) -- (9.5,8);

            \draw[black, ultra thick] (0, 4.2) -- (1.8, 4.2);
            \draw[black, ultra thick] (0, 3.8) -- (1.8, 3.8);

            \draw[black, ultra thick] (2.5, 3.8) -- (6.0, 3.8) -- (6.0, 4.2) -- (3.5, 4.2) -- cycle;

            \draw[dim] (1.9, 3.1) -- (12, 3.1) node[midway, below=4pt, font=\sffamily\large\itshape] {$L_{xd}$};
            \draw[ext] (1.9, 2.6) -- (1.9, 3.6);

            \draw[dim, dashed] (9.5, 4.2) -- (12, 4.2);
            \node[above, font=\sffamily\large\itshape] at (10.75, 4.2) {$L_{s}$};

            \draw[black, thin, dash dot] (0, 4) -- (12, 4);

            \draw[dim] (-1.2, 4) -- (-1.2, 8) node[midway, left=4pt, font=\sffamily\large\itshape] {$L_y$};
            \draw[ext] (-0.8, 4) -- (-1.5, 4);
            \draw[ext] (-0.8, 8) -- (-1.5, 8);
        \end{tikzpicture}}
        \caption{Full computational domain schematic.}
        \label{fig:domain_full}
    \end{subfigure}
    \hfill
     \begin{subfigure}[b]{0.55\textwidth}
        \centering
        \resizebox{\textwidth}{!}{%
        \begin{tikzpicture}[
            >={Latex[length=1.5mm, width=0.8mm]},
            dim/.style={black, thin, <->},
            dim_label/.style={text=black, font=\sffamily\large},
            block_fill/.style={fill=black},
            block_outline/.style={draw=black, line width=0.12cm, line join=miter},
            ext/.style={black, thin}
        ]
            \draw[black, ultra thick] (-0.5, 0.9) -- (4.0, 0.9) -- cycle;
            \draw[black, ultra thick] (-0.5, -0.2) -- (4.0, -0.2) --  cycle;

            \draw[black, ultra thick] (6.5, 0.8) -- (15.0, 0.8) -- (15.0, 1.9) -- (9, 1.9) -- cycle;

            \draw[black, thin, dash dot] (-0.5, 0.3) -- (7, 0.3);

            \draw[dim] (6.5, 0.3) -- (6.5, 0.8) node[midway, left, dim_label] {$s$};

            \draw[dim] (-0.4, 1.2) -- (4.0, 1.2) node[midway, above, dim_label] {$L_{xu}$};

            \draw[dim] (-0.6, -0.2) -- (-0.6, 0.9) node[midway, left, dim_label] {$2h$};

            \draw[dim] (4, -1.1) -- (6.5, -1.1) node[midway, below, dim_label] {$W$};
            \draw[ext] (4, -0.6) -- (4, -1.2);
            \draw[ext] (6.5, -0.6) -- (6.5, -1.2);

            \draw[dim] (9, 2.2) -- (15.0, 2.2) node[midway, above, dim_label] {$L$};
            \draw[ext] (9, 2.0) -- (9, 2.5);
            \draw[ext] (15.0, 2.0) -- (15.0, 2.5);

            \draw[dim] (15.6, 0.8) -- (15.6, 1.9) node[midway, right, dim_label] {$2h$};
            \draw[ext] (15.30, 1.9) -- (15.75, 1.9);
            \draw[ext] (15.30, 0.8) -- (15.75, 0.8);

            \coordinate (A) at (10.0, 0.8);
            \coordinate (B) at (6.5, 0.8);
            \coordinate (C) at (8.5, 1.7);
            \pic [draw=black, thick, angle radius=2cm, "$\alpha$" {anchor=center, xshift=12pt, yshift=2pt, font=\sffamily\large\bfseries}]
                {angle = A--B--C};
        \end{tikzpicture}
        }
        \caption{Zoom view of the mouthpiece geometry.}
        \label{fig:domain_zoom}
    \end{subfigure}
    \caption{Schematic of the computational domain.}
    \label{fig:domain_combined}
\end{figure}
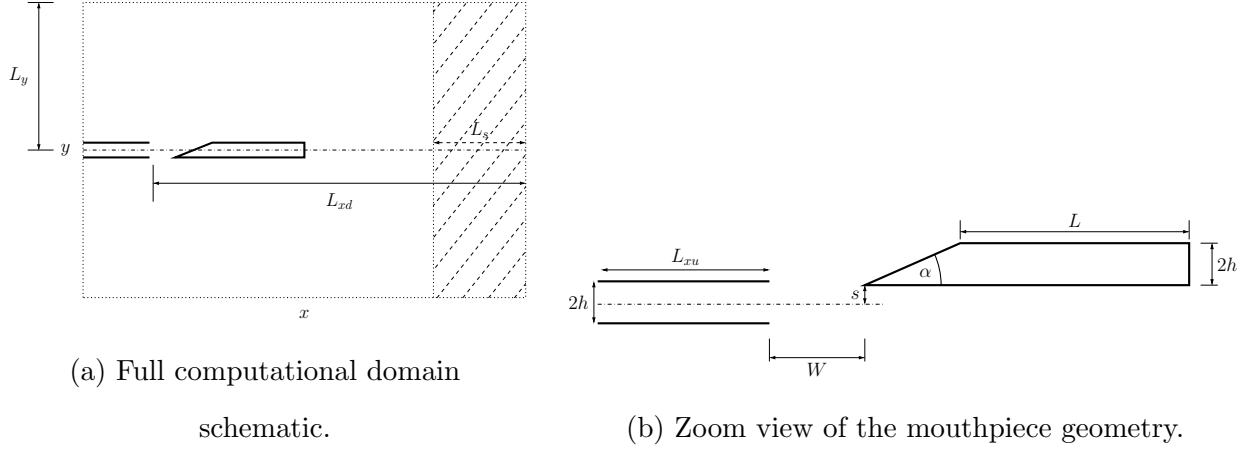

This study investigates the fundamental jet--labium interaction using an
idealized, two-dimensional recorder mouthpiece geometry. A planar jet exits a narrow
windway and interacts with an inclined sharp edge, the labium, as
illustrated in Figure~\ref{fig:domain_combined}. All lengths in this work
are non-dimensionalized using the dimensional windway height $2\hat{h}$ as
the reference length, and velocities are non-dimensionalized by the maximum (centerline) velocity $\hat{U}$ of the jet exit profile, which is $3/2$ of its mean velocity. Under this convention we define
$h = \hat{h}/2\hat{h}$, $L = \hat{L}/2\hat{h}$, $L_{xu} = \hat{L}_{xu}/2\hat{h}$,
$L_{xd} = \hat{L}_{xd}/2\hat{h}$, $W = \hat{W}/2\hat{h}$, and
$s = \hat{s}/2\hat{h}$, so that the windway height satisfies $2h = 1$ by
construction. This geometry isolates the fundamental jet--labium
interaction responsible for initiating sound in flue instruments, while
avoiding the additional complexity associated with modeling the full
resonator. Air enters through a straight channel of non-dimensional height
$2h$, traveling a streamwise distance $L_{xu}$ before exiting the
windway. Downstream of the jet exit, the labium is positioned at a
horizontal distance $W$ and forms a sharp, single-sided edge inclined at
an angle $\alpha$ relative to the jet direction. A vertical offset $s$
defines the displacement between the jet centerline and the labium tip,
allowing controlled adjustment of the jet impact location. Unlike the
idealized symmetric wedge geometries commonly employed in classical
edge-tone studies~\cite{brown1937, powell1961}, the labium considered here
is a single-sided inclined edge representative of a real recorder
mouthpiece; the configuration is therefore not geometrically symmetric
even at $s = 0$, and a non-zero offset further biases the jet
preferentially toward one side of the labium, consistent with the
geometry of a real instrument.\footnote{In the rest of this document we
have chosen to use the term ``symmetric'' for the case $s = 0$, in the
sense that there is the same vertical distance from the tip to both edges
of the windway exit.} The computational domain extends from $x = -L_{xu}$ upstream of the windway exit (at $x = 0$) to $x = L_{xd}$ downstream of it, which leaves ample room for the natural development and convection of vortical structures without interference from the outlet boundary. Its transverse half-height is defined as
\begin{equation}
    L_y = 0.5\,(L_{xd} + L_{xu}),
    \label{eq:Ly}
\end{equation}
so that the domain is square. This choice minimizes confinement effects while maintaining computational
efficiency. The Reynolds number is defined as
\begin{equation}
    Re = \frac{\hat{U}(2\hat{h})}{\hat{\nu}},
    \label{eq:Re}
\end{equation}
where $\hat{\nu}$ is the dimensional kinematic viscosity of air. The
baseline geometric parameters were set to $L = 7$, $L_{xu} = 5$, and
$L_{xd} = 50$, providing sufficient spatial separation between the inflow,
interaction region, and outlet to ensure physically realistic jet
evolution. At the upstream boundary, a parabolic Dirichlet velocity profile was
prescribed and maintained along the inlet channel, so that a fully developed
parabolic profile is established by the windway exit. The parabolic inflow
profile reflects the physical origin of the jet in a real recorder, where
viscous flow through the narrow channel naturally produces a parabolic
exit profile, rather than the uniform profile often assumed for
simplicity. Prescribing the physically representative profile also avoids
the artificial accumulation of vorticity at the jet boundaries associated
with a uniform inflow. This matters because the exit profile alters the edge-tone dynamics quantitatively: at the same Reynolds number, a parabolic jet oscillates at a $15$--$20\%$ higher Strouhal number than a top-hat jet and carries about $20\%$ more momentum flux~\cite{vaik2014a}. To prevent spurious numerical transients
at startup, the inflow profile was introduced through a smooth temporal
ramp, and the flow was seeded with a low-amplitude perturbation to
trigger the natural growth of the shear-layer instabilities that
ultimately drive vortex shedding.

The remaining boundaries were treated to minimize their influence on the
jet--labium interaction. No-slip conditions were imposed on all solid
surfaces of the labium, while a convective (zero-gradient) outflow condition
was applied at the downstream boundary to allow vortical structures to
leave the domain without generating artificial reflections. Free-slip
conditions were imposed along the transverse boundaries to limit confinement
effects and allow the jet to develop naturally, unconstrained by the lateral
extent of the domain.

Because the domain extends well beyond the downstream end of the labium,
vortices are otherwise free to impinge upon the outlet boundary and reflect
back toward the mouthpiece, a spurious phenomenon that can significantly
contaminate the physical accuracy of the results. To avoid this numerical
artifact, a sponge layer of non-dimensional length $L_s = \hat{L}_s/2\hat{h} = 20$
is introduced near the downstream boundary, as illustrated in
Figure~\ref{fig:domain_combined}(a). Within this region, an additional
damping term is added to the momentum equations that relaxes the local
velocity smoothly toward a spatially uniform target flow $(U_r, 0)$,
constant in both $x$ and $y$. The target streamwise velocity $U_r$ is set so that the flux of the incoming jet, whose mean velocity is $2/3$ of its centerline value, is spread uniformly over the full transverse extent $2L_y$ of the domain, $U_r = 2h / (3 L_y)$, so that the volumetric flux imposed at the sponge outlet matches that of the incoming jet. The strength of the damping is governed by an amplitude
$A = 0.05$, increasing smoothly from zero at the start of the sponge region
to its full value at the downstream boundary. In this way, vortical
structures are progressively attenuated as they approach the outlet,
reducing spurious reflections that would otherwise contaminate the flow
near the mouthpiece. The sensitivity of the results to the sponge-layer
parameters ($L_s$, $A$) and to the downstream domain length $L_{xd}$ was
assessed through a dedicated convergence study, presented in
Appendix~\ref{app:convergence}.

\subsection{Numerical Method}

The simulations were performed using
Direct Numerical Simulation (DNS) of the incompressible Navier--Stokes
equations. In dimensional form,
\begin{align}
    \nabla \cdot \hat{\mathbf{u}} &= 0, \\
    \frac{\partial \hat{\mathbf{u}}}{\partial \hat{t}}
    + \left(\hat{\mathbf{u}} \cdot \nabla\right)\hat{\mathbf{u}}
    &= -\nabla \hat{p} + \hat{\nu}\,\nabla^{2}\hat{\mathbf{u}},
\end{align}
where $\hat{\mathbf{u}}$ denotes the dimensional velocity vector, $\hat{p}$
the dimensional kinematic pressure, and $\hat{\nu}$ the dimensional fluid
kinematic viscosity. Non-dimensionalizing as described in
Section~\ref{subsec:geometry}, the equations actually solved take the
non-dimensional form
\begin{align}
    \nabla \cdot \mathbf{u} &= 0, \\
    \frac{\partial \mathbf{u}}{\partial t}
    + \left(\mathbf{u} \cdot \nabla\right)\mathbf{u}
    &= -\nabla p + \frac{1}{Re}\,\nabla^{2}\mathbf{u},
\end{align}
where $Re$ is the Reynolds number given by Eq.~\eqref{eq:Re}. The
incompressible assumption is well justified at the low Mach numbers
characteristic of recorder playing conditions, consistent with prior
numerical studies of the edge-tone ~\cite{vaik2014a, bamberger2004}.

An immersed boundary (IB) method~\cite{mittal2005}, based on the projection approach of Taira and Colonius~\cite{taira2007}, as implemented in the open-source solver IBMOS (Immersed Boundary Method for Optimization and Stability analysis)~\cite{fosasdepando2020ibmos}, is used to solve these equations. This approach is well suited to the present configuration, where the sharp labium edge and the narrow windway introduce strong local gradients that can be resolved without the added complexity of a body-fitted mesh.

\subsection{Control Parameters}

A systematic parametric study was conducted to investigate the influence of
geometric configuration and flow conditions on the jet--labium interaction.
Four independent control parameters were varied across physically relevant
ranges in order to capture a broad spectrum of flow behaviors.

The labium angle ($\alpha$) was selected as $15^{\circ}$, $30^{\circ}$, and
$60^{\circ}$, representing sharp to blunt edge geometries commonly
associated with variations in jet deflection and vortex formation. These
values bracket the range encountered in real flue instruments: recorders
typically feature labium angles near $15^{\circ}$, while transverse flutes
employ larger angles on the order of $60^{\circ}$, with corresponding
differences in oscillation stability and tonal character~\cite{dequand2003}.

The Reynolds number ($Re$), defined in Eq.~\eqref{eq:Re}, was varied
between 200 and 700 ($Re = 200,\, 330,\, 500,\, 700$) to span regimes from
weakly unstable shear layers to strongly nonlinear oscillatory flows. This
definition is based on the windway geometry rather than the labium
dimensions, consistently with the convention used in edge-tone studies
where the corresponding length scale (referred to there as the nozzle
height) is used~\cite{vaik2014a,bamberger2004}.

The vertical offset ($s$), defined as the displacement between the jet
centerline and the labium tip, positive when the tip lies above the jet
centerline (Figure~\ref{fig:domain_combined}(b)), was examined over the range
$s = -0.5,\,-0.2,\,-0.1,\,0,\,0.1,\,0.2,\,0.5$. Both positive and negative offsets were
considered in order to evaluate symmetry-breaking effects and their role in
directing the jet toward or away from the edge.

The mouth distance ($W$), corresponding to the horizontal separation between the jet exit
and the labium tip, was varied between $3$ and $12$ ($W = 3, 4, 5, 6, 6.5, 7, 8$ and $12$). This parameter
directly influences the jet travel time prior to impact and is therefore
expected to affect vortex roll-up and oscillation characteristics.
This parameter is found in other edge-tone studies as nozzle-to-edge
distance, and it has been shown to govern the oscillation stage and the
Strouhal number of the resulting flow, defined here as
\begin{equation}
    St = \frac{f (2h)}{U},
    \label{eq:St}
\end{equation}
where $f$ is the oscillation frequency and $U = 1$ the non-dimensional centerline velocity of the jet exit profile~\cite{brown1937,vaik2014a,vaik2014b}. When several discrete spectral peaks are present in the force spectrum, $St_1$ denotes the fundamental of the Stage I oscillation, the dominant peak at the baseline $W = 4$ ($St_1 \approx 0.07$), and $St_2, \dots$ denote the additional peaks.

The parameters were varied around the baseline configuration ($W = 4$, $\alpha = 15^{\circ}$, $s = 0$, $Re = 330$), and more than $200$ simulations were performed in total. This structured exploration of the parameter space enables
the identification of governing flow regimes and provides a consistent basis
for assessing how geometric and flow variations control vortex dynamics and
unsteady jet behavior.

\subsection{Spatiotemporal Resolution}
\label{subsec:resolution}

To resolve the strong local gradients near the jet exit and the labium
tip, the computational domain employed a stretched Cartesian mesh, with
enhanced resolution concentrated in this region, where velocity gradients
are steepest and vortex formation first occurs. Near the jet exit and the
labium tip, a fine grid spacing of $\Delta x = \Delta y = 1/n_1$ (with
$n_1 = 24$) was used, while a coarser far-field spacing of $1/n_2$ was
applied away from this region, with the mesh smoothly stretched between
the two: for the baseline parametric sweep ($Re \le 500$), a far-field
grid spacing of $1/n_2$ (with $n_2 = 6$) was used; for the highest
Reynolds number cases ($Re = 700$), this far-field resolution was refined
to $n_2 = 12$ to adequately resolve the sharper velocity gradients
associated with the thinner, higher-$Re$ shear layers. For the baseline
configuration ($W = 4$, $\alpha = 15^\circ$, $Re = 330$), this yields a
total mesh of $701 \times 403$ points (282{,}503 points). The time step
was not fixed a priori but was instead chosen adaptively so that the
Courant number based on the near-field grid spacing did not exceed
a target value of $0.4$, keeping the numerical scheme stable relative to
the finest resolved scales of the flow. Each simulation was advanced for
a non-dimensional duration of $T_{\mathrm{sim}} = 2000$, a duration sufficient to allow
the initial transient to decay and the flow to reach a statistically
saturated, self-sustained oscillatory regime, from which the diagnostic
quantities reported in Section~\ref{sec:results} were extracted. The
sensitivity of the results to this spatial resolution is assessed in
Appendix~\ref{app:convergence}.

\subsection{Spectral Analysis}
\label{subsec:spectral_analysis}

Spectra are computed with Welch's method. The first $20\%$ of each record, which contains the initial transient, is discarded, so that the analyzed record covers $1600$ non-dimensional time units, and the mean of the remaining signal is removed. The record is divided into segments of one eighth of its length ($200$ time units, giving a frequency resolution $\Delta St = 0.005$) with $50\%$ overlap, which gives $15$ segments. Each segment is multiplied by a Hann window, and the squared moduli of the Fourier transforms of the segments are averaged, which reduces the random scatter of a single Fourier transform at the cost of frequency resolution. The result is scaled so that the height of a spectral line equals the mean-square value of the corresponding oscillation. The figures show its square root, the root-mean-square (RMS) amplitude of the oscillation in each frequency bin, in the units of the analyzed signal; this equals $1/\sqrt{2}$ times the amplitude of a sinusoid of the same frequency. The zero-frequency bin is not shown. The spectra are not normalized, except in Figure~\ref{fig:flow_visualization_baseline}, where each signal is divided by its maximum. Where peak frequencies are quoted in Section~\ref{subsec:effect_mouth_distance} and Table~\ref{tab:W_groups}, they are obtained from spectra computed with segments of half the record length ($\Delta St = 0.00125$, three segments) and a parabolic interpolation of the peak, which separate the closely spaced peaks at larger $W$ better than the spectra plotted in the figures. Cases with strong amplitude modulation ($W \ge 7$ and $|s| = 0.5$) are not statistically stationary over the analyzed record, so their spectral amplitudes are averages over it and are indicative.

\section{Results and Discussion}\label{sec:results}

Before mapping the full parameter space, it is useful to first describe the
qualitative behavior of a characteristic case, in order to illustrate the
underlying flow physics that motivate the parametric study which follows.
The self-sustained oscillations of a planar jet impinging on a sharp labium
wedge arise from the interaction between downstream hydrodynamic
instability growth and upstream hydrodynamic
feedback~\cite{powell1961, nagy2021}. To systematically map out how the
localized mouthpiece boundaries dictate this feedback loop independently of
a downstream acoustic resonator, the parametric sweep described in Section~\ref{sec:methodology} is
analyzed here first in terms of the Reynolds number $Re$, which governs
the initial growth of the shear-layer instability. The three
geometric control parameters then follow in turn. First, the mouth distance
$W$, which determines how far shear-layer disturbances must travel from the jet
exit before reaching the labium tip and, as a consequence,
which discrete oscillation stage is selected and whether multiple stages
compete simultaneously. Next, the vertical alignment offset $s$, which breaks
the geometric symmetry of the jet--labium interaction. Finally, the
labium wedge inclination angle $\alpha$, which locally redirects and
deflects the impinging shear layer at the tip of the wedge without
altering the underlying instability mechanism.

\subsection{Flow Regimes and Onset of Oscillation}
\label{subsec:overview_regimes}

Before examining each geometric parameter individually, we first
look at the primary flow regimes across Reynolds numbers and confirm that
our diagnostic signal choice is reliable. Only four Reynolds numbers were simulated ($Re = 200$, $330$, $500$ and $700$), so the critical Reynolds number for the onset of oscillation is not determined here.

At $Re = 200$ (not shown), self-sustained edge-tone s depend on the vertical alignment offset $s$. Under symmetric conditions ($s = 0$) and for the near-centerline offsets $s = \pm 0.1$ and $\pm 0.2$, a stable edge-tone oscillation is maintained, whereas at $s = \pm 0.5$ the oscillation is almost completely suppressed.

\begin{figure}[!htbp]
    \centering
    \begin{subfigure}[b]{0.24\textwidth}
        \centering
        \includegraphics[width=\textwidth]{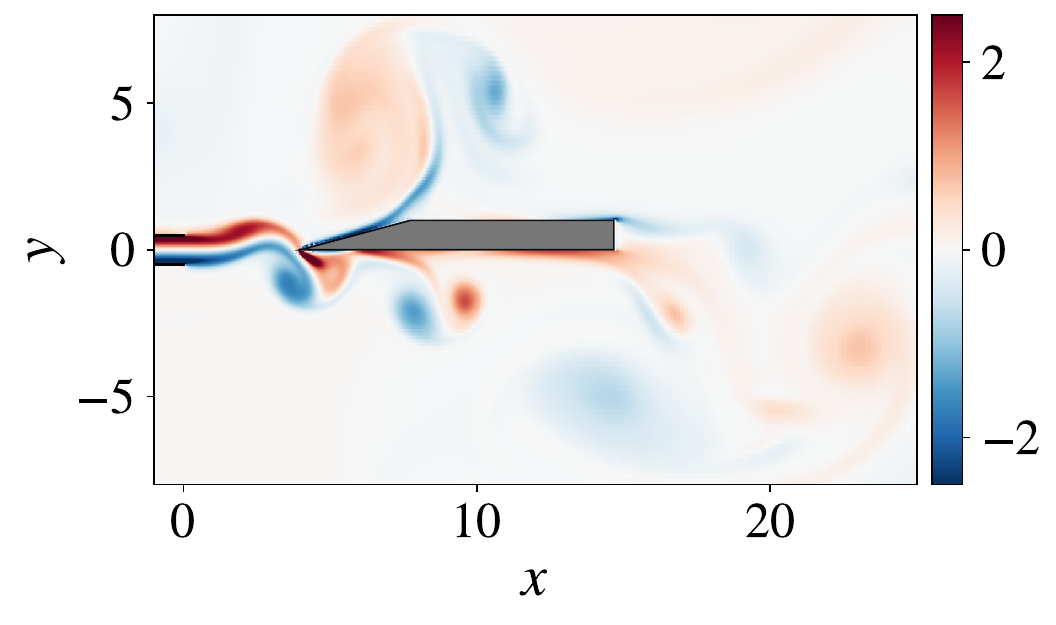}
        \caption{$t = 1850$}
        \label{fig:vort_1850}
    \end{subfigure}
    \hfill
    \begin{subfigure}[b]{0.24\textwidth}
        \centering
        \includegraphics[width=\textwidth]{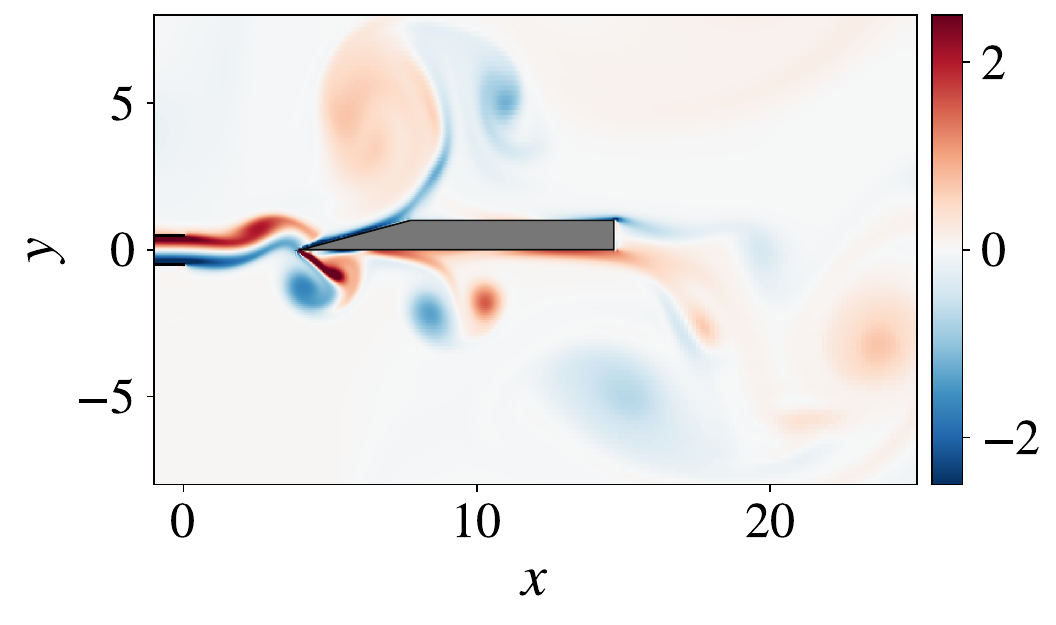}
        \caption{$t = 1852$}
        \label{fig:vort_1852}
    \end{subfigure}
    \hfill
    \begin{subfigure}[b]{0.24\textwidth}
        \centering
        \includegraphics[width=\textwidth]{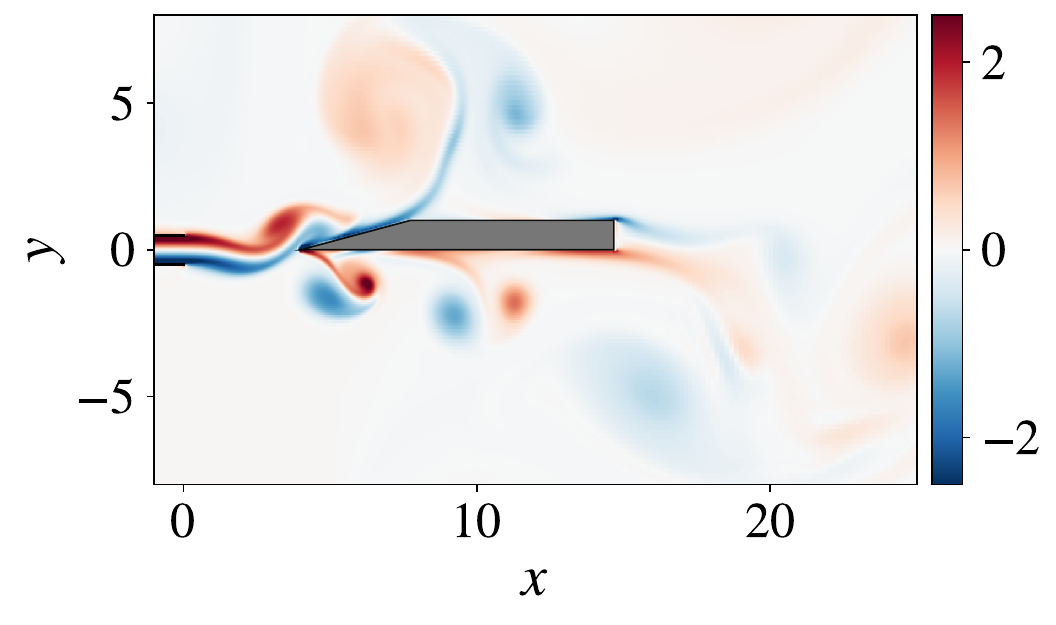}
        \caption{$t = 1855$}
        \label{fig:vort_1855}
    \end{subfigure}

    \vspace{0.3cm}

    \begin{subfigure}[b]{0.24\textwidth}
        \centering
        \includegraphics[width=\textwidth]{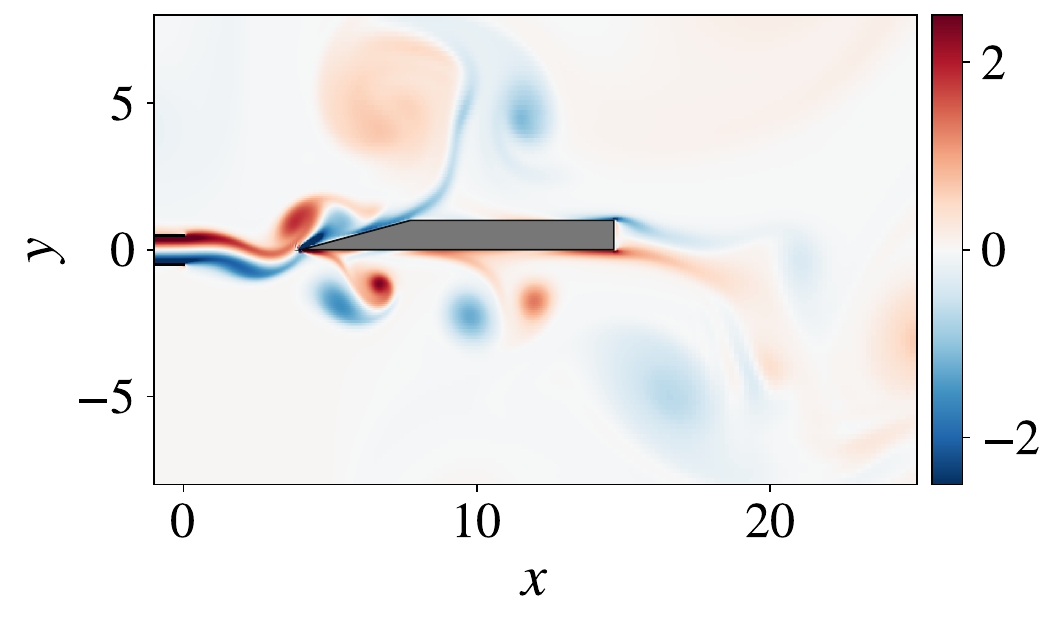}
        \caption{$t = 1857$}
        \label{fig:vort_1857}
    \end{subfigure}
    \hfill
    \begin{subfigure}[b]{0.24\textwidth}
        \centering
        \includegraphics[width=\textwidth]{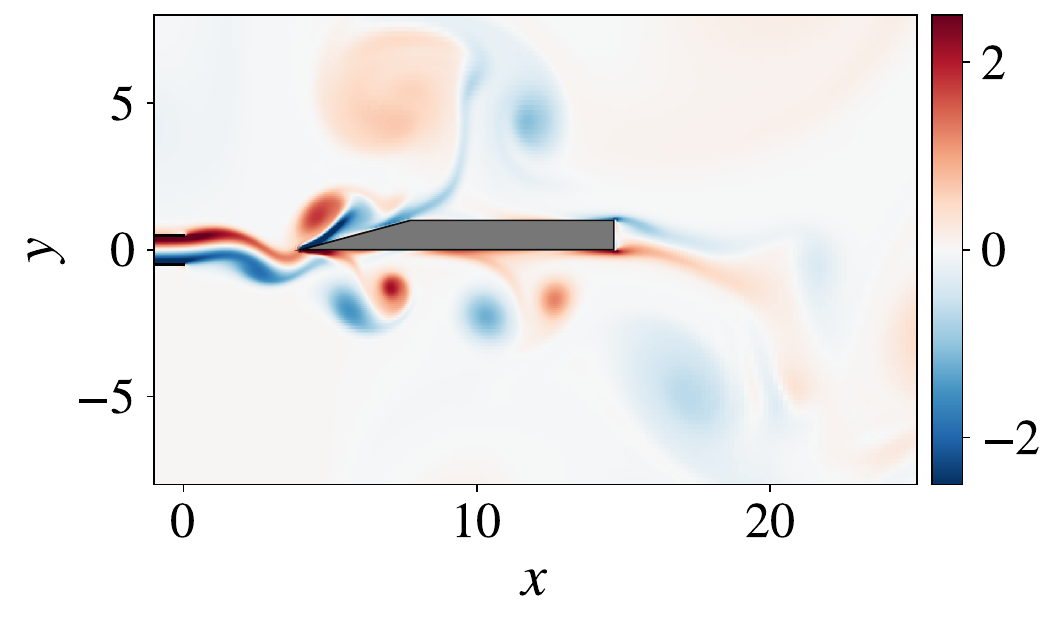}
        \caption{$t = 1859$}
        \label{fig:vort_1859}
    \end{subfigure}
    \hfill
    \begin{subfigure}[b]{0.24\textwidth}
        \centering
        \includegraphics[width=\textwidth]{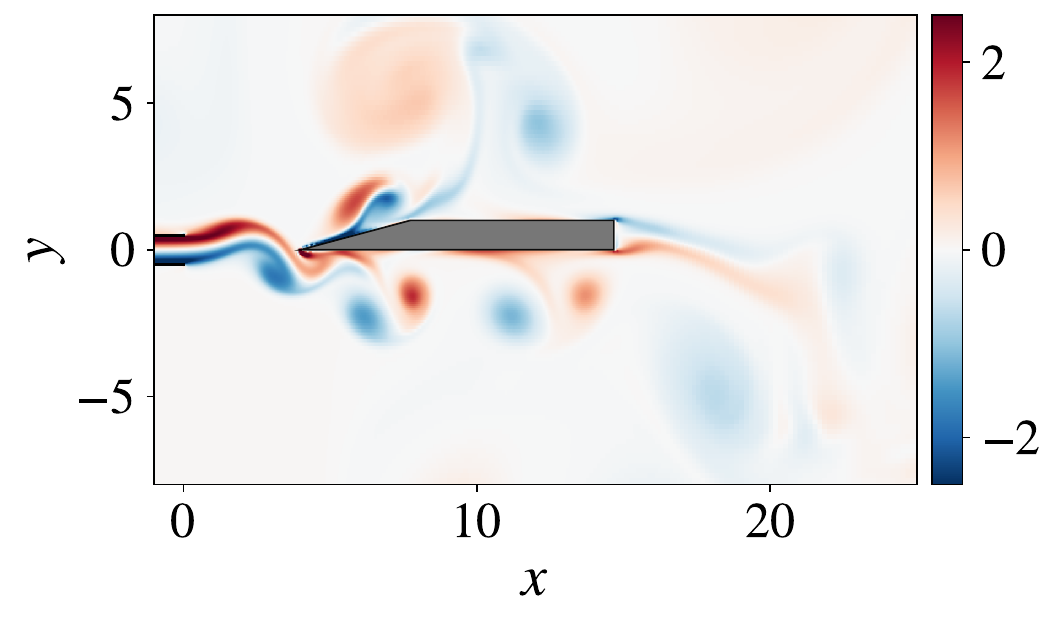}
        \caption{$t = 1862$}
        \label{fig:vort_1862}
    \end{subfigure}

    \vspace{0.3cm}

    \begin{subfigure}[t]{0.38\textwidth}
        \centering
        \includegraphics[width=\textwidth]{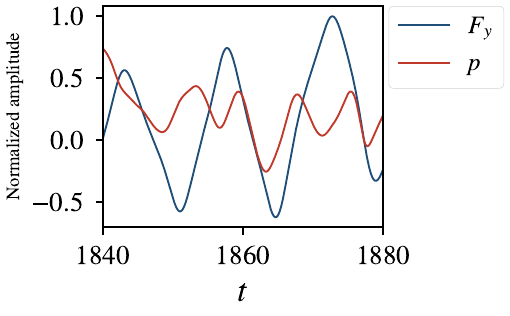}
        \caption{Time histories of $F_y$ and $p$}
        \label{fig:Fy_p_timehistory}
    \end{subfigure}
    \hfill
    \begin{subfigure}[t]{0.38\textwidth}
        \centering
        \includegraphics[width=\textwidth]{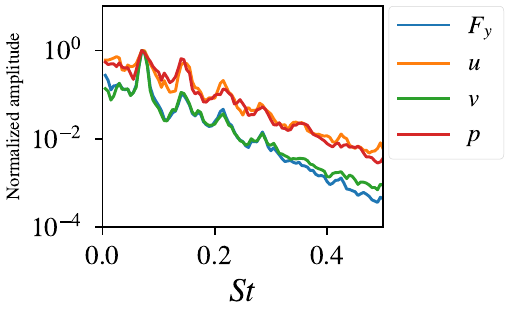}
        \caption{Overlaid Welch amplitude spectrum ($F_y$, $u$, $v$, $p$)}
        \label{fig:validation_fft}
    \end{subfigure}

    \caption{Oscillatory dynamics of the jet at $Re = 330$ ($W = 4$, $\alpha = 15^{\circ}$, $s = 0$). (a--f) Instantaneous vorticity fields, zoomed to the windway exit, jet, and labium region, across six sequential non-dimensional times spanning most of one oscillation period ($T \approx 1/St_1 \approx 14.3$, with $St_1 \approx 0.07$), illustrating the cycle of shear-layer roll-up, vortex advection, and tip-splitting at the labium that sustains the periodic Stage I edge-tone . (g) Time histories of $F_y$ and of the pressure $p$ from the same near-labium probe (each normalized by its maximum over the whole analyzed record), showing that both signals oscillate with the same period. (h) Overlaid Welch amplitude spectra of the transverse labium force $F_y$ and of the velocity components $u$, $v$ and pressure $p$ sampled near the labium (each normalized by its maximum over the whole analyzed record). All four signals have a peak at the same frequency, $St_1 \approx 0.07$, so the periodicity seen in (a--f) is not an artifact of the particular signal monitored.}
    \label{fig:flow_visualization_baseline}
\end{figure}

At $Re = 330$, the flow transitions into a robust, fully developed oscillatory regime. Figure~\ref{fig:flow_visualization_baseline} illustrates this regime at baseline conditions ($W = 4$, $\alpha = 15^{\circ}$, $s = 0$) through six sequential vorticity snapshots spanning most of one oscillation period, together with time histories of $F_y$ and the pressure $p$ and the Welch amplitude spectra\footnote{Spectra are computed as described in Section~\ref{subsec:spectral_analysis}.} of the transverse labium force $F_y$ and, for comparison, of the velocity components $u$, $v$ and the pressure $p$ sampled near the labium.

Discrete vortical structures of alternating sign develop along the upper and lower shear layers, advecting downstream and splitting at the labium tip to sustain the classic hydrodynamic loop~\cite{powell1961,holger1977}. The time histories in Figure~\ref{fig:flow_visualization_baseline}(g) and the spectra in Figure~\ref{fig:flow_visualization_baseline}(h) show that $F_y$, $u$, $v$ and $p$ oscillate at the same fundamental frequency, $St_1 \approx 0.07$, agreeing to within the frequency resolution of the Welch estimate. This agreement shows that the flow oscillates at a single dominant frequency~\cite{rayleigh1896theory}. Consequently, the transverse force $F_y$ on the labium, a direct, single-point surface-integrated measure of the hydrodynamic loading that drives the jet--labium feedback loop and straightforward to extract consistently across all simulated cases, is used as a proxy for the flow behavior, which is correlated with $u$, $v$ and $p$, and is selected as the primary diagnostic variable for all subsequent parametric evaluations.

\begin{figure}[!htbp]
    \centering
    \begin{subfigure}[b]{0.22\textwidth}
        \centering
        \includegraphics[width=\textwidth]{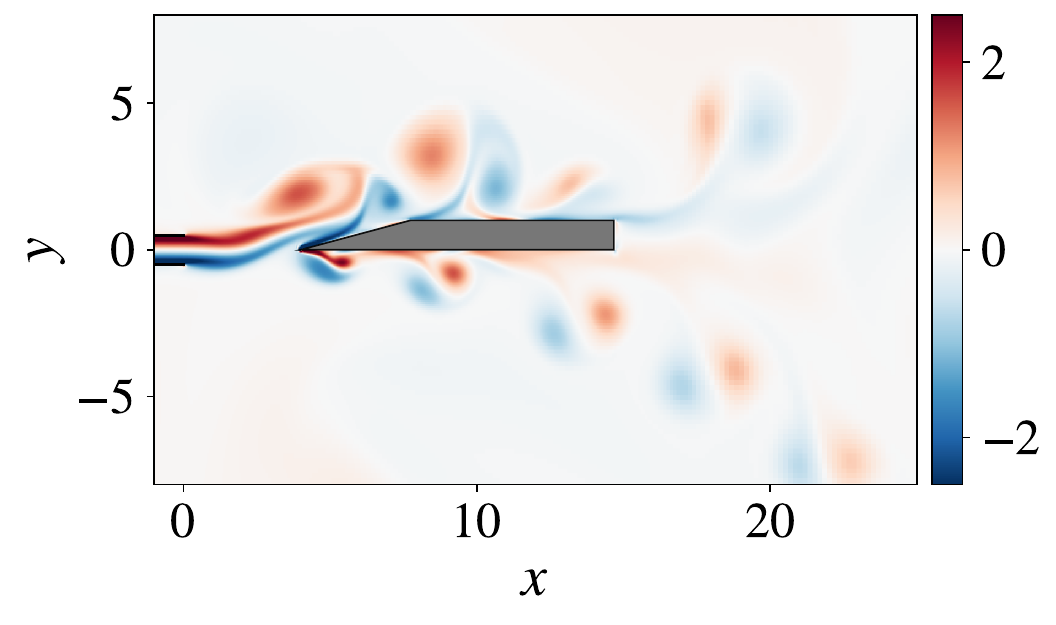}
        \caption{$t = 985$}
        \label{fig:defl_up_1}
    \end{subfigure}
    \hfill
    \begin{subfigure}[b]{0.22\textwidth}
        \centering
        \includegraphics[width=\textwidth]{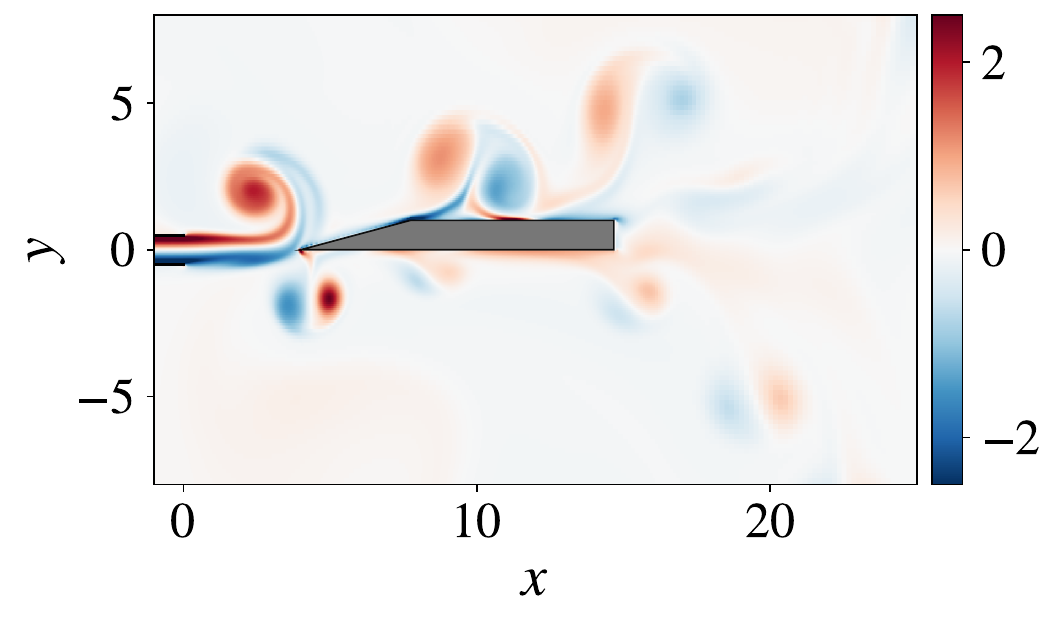}
        \caption{$t = 1006$}
        \label{fig:defl_up_2}
    \end{subfigure}
    \hfill
    \begin{subfigure}[b]{0.22\textwidth}
        \centering
        \includegraphics[width=\textwidth]{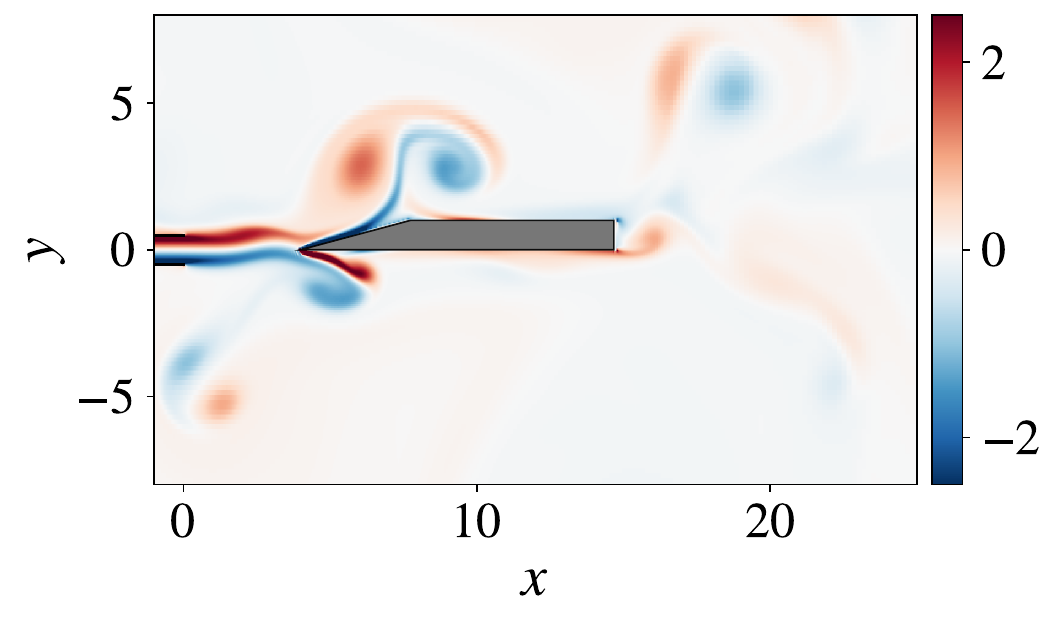}
        \caption{$t = 1030$}
        \label{fig:defl_up_3}
    \end{subfigure}

    \vspace{0.15cm}

    \begin{subfigure}[b]{0.22\textwidth}
        \centering
        \includegraphics[width=\textwidth]{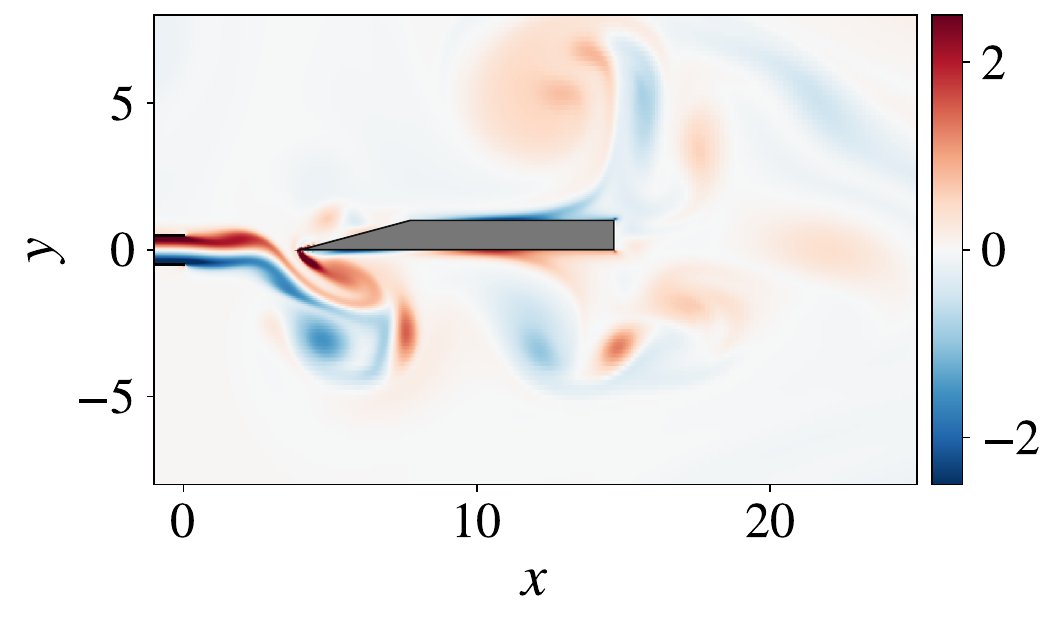}
        \caption{$t = 1920$}
        \label{fig:defl_down_1}
    \end{subfigure}
    \hfill
    \begin{subfigure}[b]{0.22\textwidth}
        \centering
        \includegraphics[width=\textwidth]{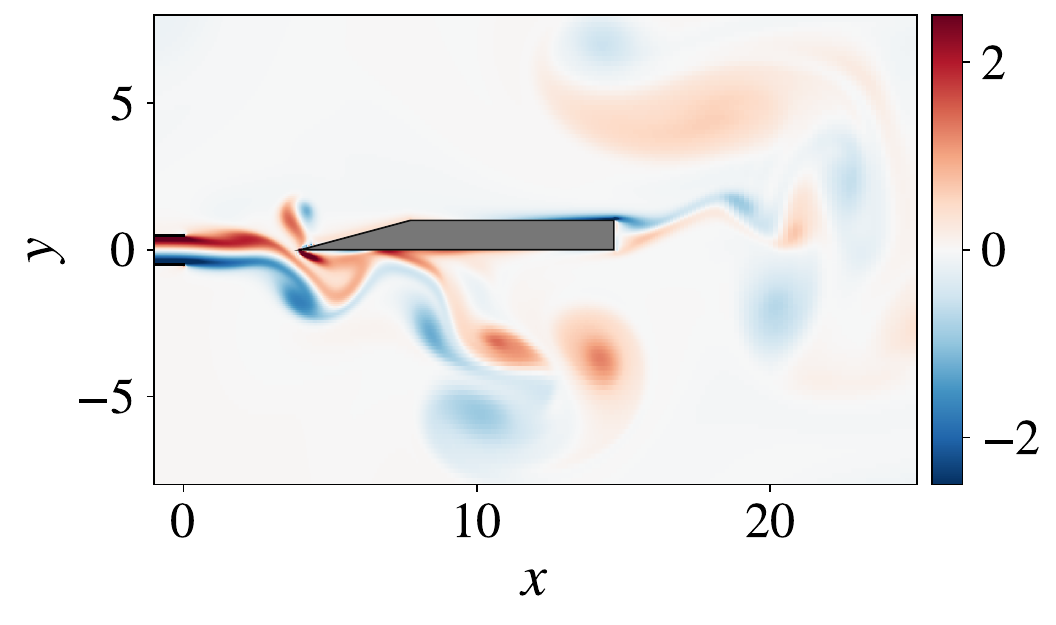}
        \caption{$t = 1949$}
        \label{fig:defl_down_2}
    \end{subfigure}
    \hfill
    \begin{subfigure}[b]{0.22\textwidth}
        \centering
        \includegraphics[width=\textwidth]{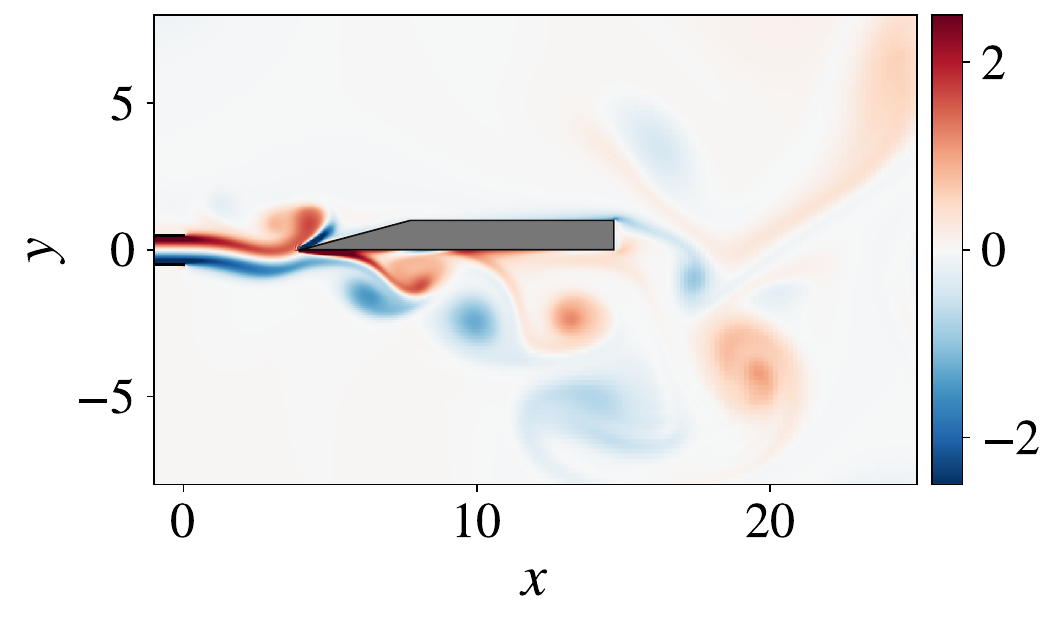}
        \caption{$t = 1970$}
        \label{fig:defl_down_3}
    \end{subfigure}

    \vspace{0.15cm}

    \begin{subfigure}[b]{0.72\textwidth}
        \centering
        \includegraphics[width=\textwidth]{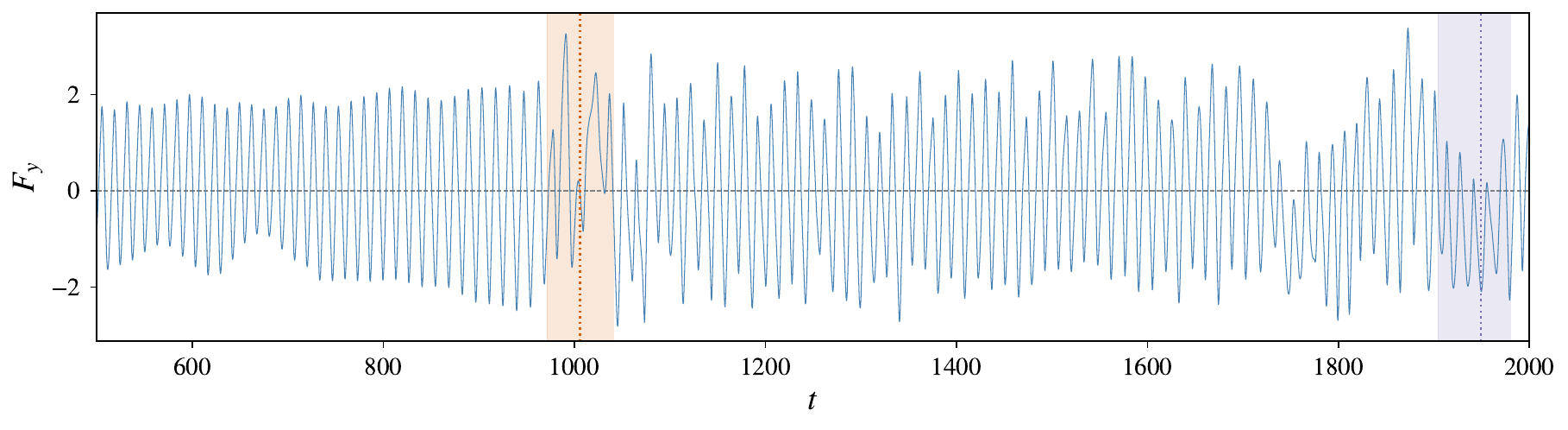}
        \caption{Time history of $F_y$}
        \label{fig:intermittency_timehistory}
    \end{subfigure}

    \caption{Intermittency of the baseline oscillation at $Re = 330$ ($W = 4$, $\alpha = 15^{\circ}$, $s = 0$): two episodes in which the jet stays biased to one side of the labium for several cycles before returning to symmetric shedding. (a--c) Upward-biased episode ($t \in [971, 1041]$) at onset ($t = 985$), maximum bias ($t = 1006$) and recovery ($t = 1030$). (d--f) Downward-biased episode ($t \in [1904, 1980]$) at $t = 1920$, $1949$ and $1970$. (g) Time history of $F_y$ over the saturated record, with the upward (orange) and downward (purple) episodes shaded and their peak-bias times marked by dotted lines.}
    \label{fig:intermittency}
\end{figure}

The baseline case is periodic over most of the simulated record, but the time history of $F_y$ also exhibits intervals during which the jet deflects toward one side of the labium and remains asymmetrically biased for several oscillation cycles before returning to symmetric shedding. Bias intervals were identified from a rolling-window average of the saturated $F_y(t)$ record (window length $40$ non-dimensional time units, exceeding one oscillation period), which isolates sustained deflections from the underlying periodic oscillation. Figure~\ref{fig:intermittency} illustrates two such episodes, one with the jet biased upward and one with the jet biased downward, each traced across three snapshots showing the onset, peak, and recovery of the deflection, together with the full time history of $F_y$ over the saturated record. The one-sided vortex organization at the peak-bias frames contrasts with the balanced, alternating shedding of the clean periodic cycle in Figure~\ref{fig:flow_visualization_baseline}. Two representative episodes, selected from several occurrences, are shown; their frequency and duration are not quantified; a similar intermittency is seen in the time histories at nonzero offsets (Section~\ref{subsec:influence_offset}).

\subsection{Dependence on Reynolds Number}
\label{subsec:dependence_re}

Having outlined the threshold conditions at low Reynolds numbers, we examine how the flow evolution scales as the Reynolds number is systematically increased across $Re = 330, 500,$ and $700$ under symmetric baseline conditions ($s = 0$). At $Re = 200$ the oscillation is sustained only for near-centerline offsets ($|s| \le 0.2$) and is almost completely suppressed at $|s| = 0.5$ (Section~\ref{subsec:overview_regimes}), whereas for higher Reynolds numbers ($Re \ge 330$) an oscillation is present for all offsets, although it is strongly reduced at $|s| = 0.5$.

Here, we define the oscillatory regime as the flow state characterized by self-sustained, periodic limit-cycle oscillations exhibiting a sharp, dominant spectral peak in the transverse force $F_y$ on the labium. Within this oscillatory regime, the non-dimensional frequency quantified
by the fundamental Strouhal number $St$ (Eq.~\eqref{eq:St}) remains nearly
constant across $Re = 330, 500,$ and $700$. This suggests that the global flow dynamics and the convective transit
time of shear-layer disturbances between the jet exit and the labium do
not significantly depend on the Reynolds number, and are governed primarily by
the mouth geometry rather than by viscosity. This behavior is fully consistent with classical edge-tone scaling laws, in which the saturated Strouhal number depends primarily on the non-dimensional mouth distance $W$ and converges asymptotically to a Reynolds-number-independent plateau at moderate-to-high $Re$ regimes~\cite{vaik2014b}; Bamberger \textit{et al.}~\cite{bamberger2004} likewise observed only a weak dependence on the Reynolds number.

\begin{figure}[!htbp]
    \centering
    \includegraphics[width=0.6\textwidth]{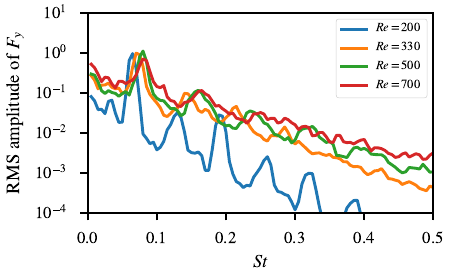}
    \caption{Welch amplitude spectra of the transverse force $F_y$ on the labium wedge across varying Reynolds numbers ($Re = 200$, $330$, $500$, $700$) under symmetric conditions ($W = 4$, $\alpha = 15^{\circ}$, $s = 0$). All cases share the primary peak at $St \approx 0.07$. The $Re = 200$ spectrum decays much faster at higher $St$, whereas the higher Reynolds numbers ($Re \ge 330$) retain more energy in the harmonics, which increases with $Re$.}
    \label{fig:fft_dependence_re}
\end{figure}

The amplitude of the hydrodynamic force fluctuations on the labium tip generally
increases with the Reynolds number from $Re=200$ to $Re=500$, reflecting the
higher dynamic pressure and enhanced kinetic energy flux of the jet at higher
inertia; however, the peak amplitude at $Re=700$ is slightly lower than at
$Re=500$, indicating that this trend is not strictly monotonic. Since the far-field resolution differs between $Re = 500$ and $Re = 700$ ($n_2 = 6$ and $12$), this difference may be affected by the mesh and is not interpreted further. A key observation from the spectral signatures is that the underlying structure of the oscillation, more specifically the total number of spectral peaks, their harmonic spacing, and their relative energy distribution, is largely preserved across the Reynolds numbers studied, although the energy in the harmonics increases with $Re$ (Figure~\ref{fig:fft_dependence_re}). This consistency indicates that, within this parameter envelope, the flow remains within Stage I. We define Stage I, following the phase-locking framework of Holger \textit{et al.}~\cite{holger1977}, as the fundamental edge-tone regime in which the windway-to-labium distance satisfies the lowest-order geometric condition for constructive feedback between the convecting vortex street and the jet origin, producing a single dominant spectral peak together with its integer harmonics rather than competing, incommensurate frequencies. We refer to the geometric condition governing this selection as the
\emph{phase-matching condition} throughout the remainder of this work. Consequently, variation of $Re$ within the active regime appears to modulate primarily the nonlinear saturation amplitude of the flapping jet, with no transition to higher-order hydrodynamic stages observed within this range.

Because the fundamental Strouhal number and flow structure remain essentially unchanged across $Re \in [330, 700]$, $Re = 330$ is selected as the baseline operating condition for the broader geometric parameter sweeps. It is the lowest Reynolds number tested at which an oscillation is present for every offset considered, although much weaker at $|s| = 0.5$, whereas at $Re = 200$ it is almost completely suppressed there; it therefore provides a well-resolved and representative baseline for mapping the effects of mouth geometry.

\subsection{Effect of Mouth Distance}
\label{subsec:effect_mouth_distance}

The mouth distance $W$ strongly influences the selection, development, and stability of the hydrodynamic oscillation regime. Its role is closely analogous to that of the nozzle-to-edge distance in classical edge-tone configurations~\cite{brown1937,vaik2014a,vaik2014b}, where this geometric parameter serves as a convective time-delay line, regulating the travel time of jet shear-layer disturbances relative to the phase of the upstream pressure response, and thereby functioning as the main parameter that selects the feedback stage.

At the baseline mouth distance $W = 4$, the fluid jet travels a relatively short distance before impinging upon the sharp labium edge. The force spectrum is dominated by a single sharp fundamental peak, at $St_1 \approx 0.072$, and a second harmonic appears at $\approx 2St_1$. This integer harmonic ratio is characteristic of a single-stage oscillation (Stage I)~\cite{brown1937}. In this regime the edge-tone  is stable and repeatable, with the force dominated by the fundamental. The case $W = 3$ (not shown) is also dominated by a single fundamental, at $St \approx 0.092$.

The baseline fundamental, $St_1 \approx 0.07$ at $W = 4$, is the reference frequency of this study. As $W$ increases, the Stage I peak moves to lower $St$ in such a way that $St \cdot (W/2h)$ stays close to $0.29$: it equals $0.286$, $0.291$, $0.286$, $0.296$ and $0.295$ for $W = 4$, $5$, $6$, $6.5$ and $7$ (Table~\ref{tab:W_groups}). Its frequency therefore scales as $1/W$, consistent with the exponent of $1$ reported in classical edge-tone studies~\cite{vaik2014a,bamberger2004}, and the amplitude of the spectrum near $St = 0.07$ falls to between $7\%$ and $34\%$ of the strongest peak for $W \ge 5$. Alongside the Stage I peak, a second stage appears as peaks at lower frequencies, near $St \cdot (W/2h) \approx 0.2$, and grows gradually with $W$. Its amplitude relative to the Stage I peak is about $9\%$ at $W = 4$, and it is only weakly present at $W = 5$ and $6$ ($19$--$20\%$). It reaches $43$--$46\%$ at $W = 6.5$ and $7$, where it is clearly established while the Stage I peak still has the highest amplitude, and it exceeds the Stage I peak at $W = 8$.

\begin{table}[!htbp]
\centering
\begin{tabular}{cccc}
\hline
$W$ & Dominant $St$ & $St \cdot (W/2h)$ & Second stage / Stage I \\
\hline
4   & 0.0715 & 0.286 & 0.09 \\
5   & 0.0582 & 0.291 & 0.19 \\
6   & 0.0476 & 0.286 & 0.20 \\
6.5 & 0.0456 & 0.296 & 0.43 \\
7   & 0.0422 & 0.295 & 0.46 \\
8   & 0.0249 & 0.200 & 1.18 \\
12  & 0.0161 & 0.193 & --- \\
\hline
\end{tabular}
\caption{Dominant peak of the force spectrum and amplitude of the second-stage peaks relative to the Stage I peak, at $Re = 330$, $\alpha = 15^{\circ}$, $s = 0$. The second stage is evaluated at $0.15 \le St \cdot (W/2h) \le 0.23$ and the Stage I peak at $0.25 \le St \cdot (W/2h) \le 0.33$, from spectra computed with segments of half the record length (Section~\ref{subsec:spectral_analysis}). At $W = 8$ the ratio exceeds $1$ because the second stage has become the dominant one; at $W = 12$ the two stages are no longer distinct and the ratio is not given.}
\label{tab:W_groups}
\end{table}

At $W = 8$ the strongest peak lies at $St \approx 0.025$ ($St \cdot (W/2h) \approx 0.20$) and is accompanied by closely spaced second-stage peaks of similar amplitude between $St \approx 0.018$ and $0.028$, while a peak at the Stage I value, at $St \approx 0.036$, is still present with about $85\%$ of the amplitude of the strongest one. The onset of the second stage is therefore gradual: it is only weakly present at $W = 5$ and $6$ and clearly established from $W = 6.5$. This is consistent with the phase-based stage-selection criterion of Holger \textit{et al.}~\cite{holger1977}, in which the discrete edge-tone stages correspond to street wavelengths satisfying an in-phase feedback condition between the vortex street and the jet origin; stage transitions in their model are further associated with comparable growth-rate amplification between neighboring stages. The present results do not independently verify either mechanism but are consistent with the associated transition location.

The second stage lies below the Stage I peak in $St \cdot (W/2h)$ ($\approx 0.2$ against $0.29$), whereas the higher stages of the classical edge-tone ladder lie above the first stage at a given distance and speed~\cite{brown1937,vaik2014a}. We therefore use ``second stage'' only to denote a second, competing oscillation of the feedback loop, and do not identify it with the higher stages of Brown. Its nature (a modulation or subharmonic of Stage I, or a wake or labium mode) is not determined here.

\begin{figure}[!htbp]
    \centering
    \begin{subfigure}[b]{0.32\textwidth}
        \centering
        \includegraphics[width=\textwidth]{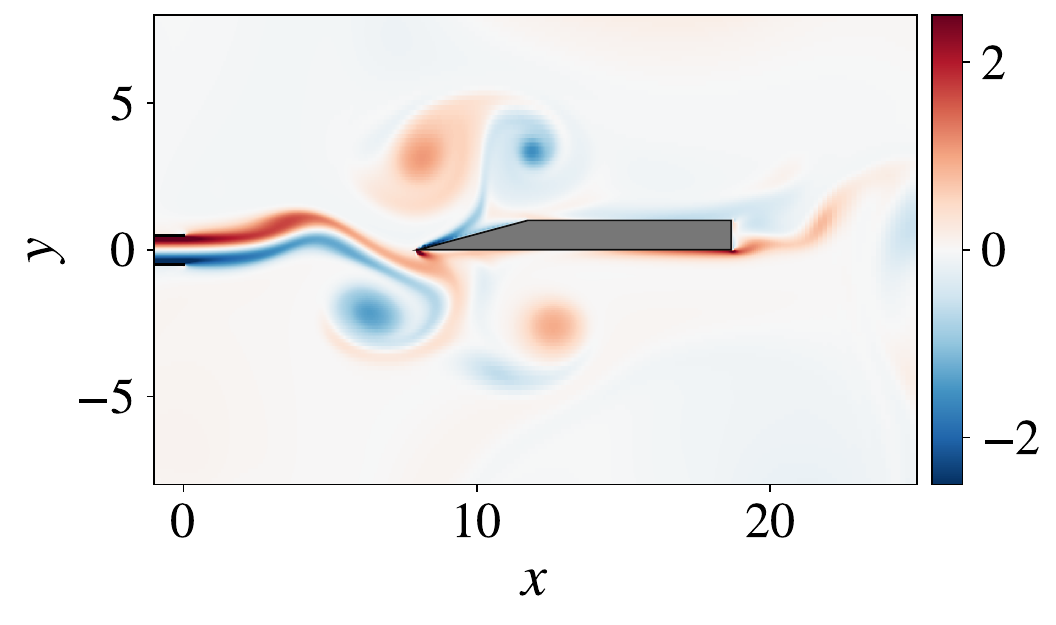}
        \caption{$t = 1885$}
        \label{fig:w8_t1}
    \end{subfigure}
    \hfill
    \begin{subfigure}[b]{0.32\textwidth}
        \centering
        \includegraphics[width=\textwidth]{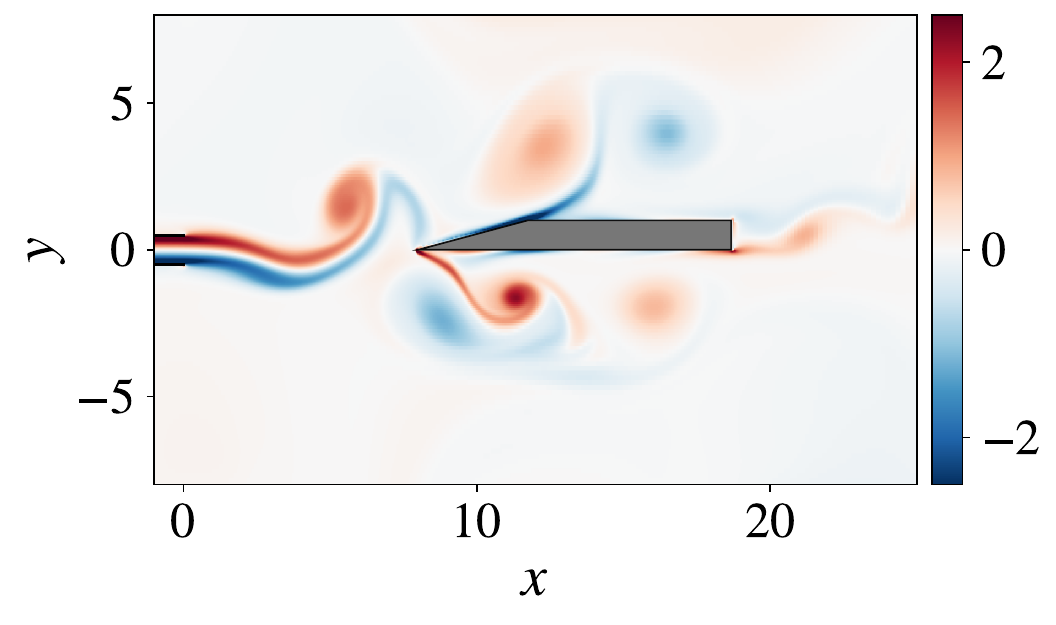}
        \caption{$t = 1899$}
        \label{fig:w8_t2}
    \end{subfigure}
    \hfill
    \begin{subfigure}[b]{0.32\textwidth}
        \centering
        \includegraphics[width=\textwidth]{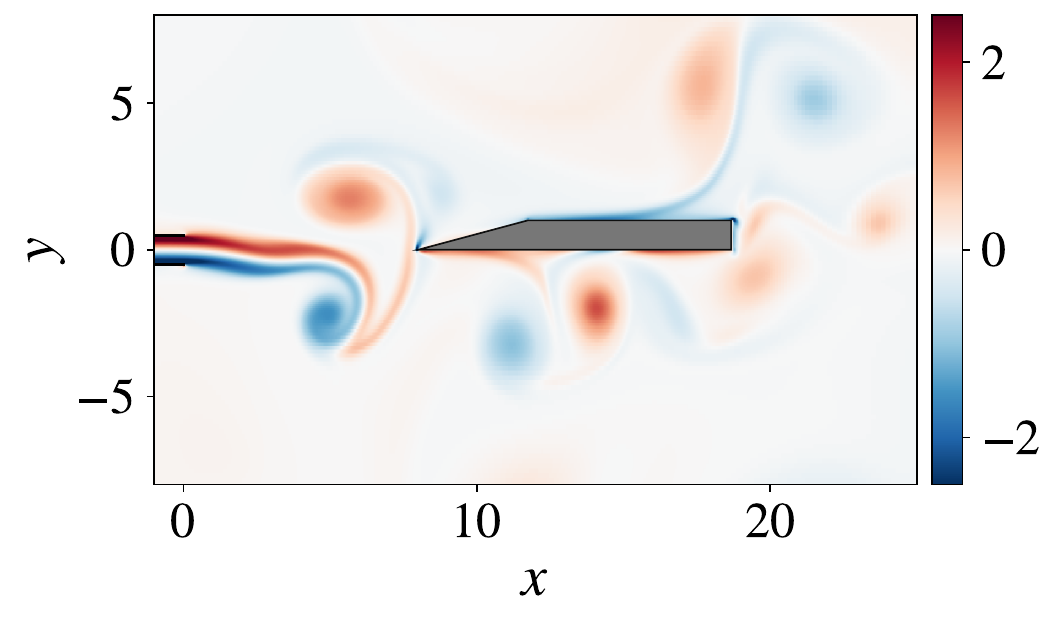}
        \caption{$t = 1912$}
        \label{fig:w8_t3}
    \end{subfigure}

    \vspace{0.35cm}

    \begin{subfigure}[b]{0.32\textwidth}
        \centering
        \includegraphics[width=\textwidth]{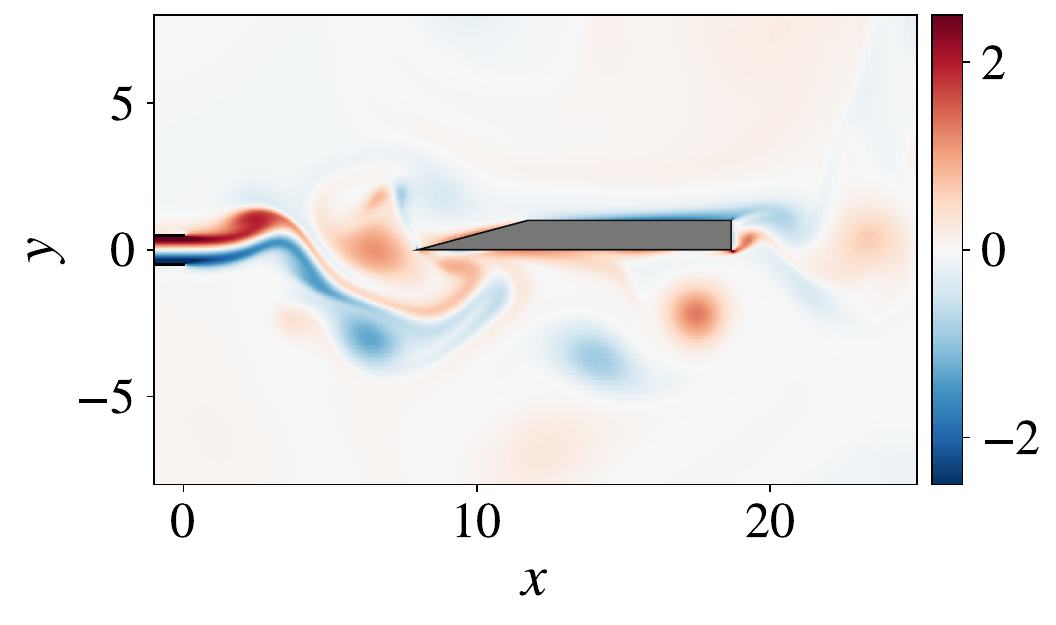}
        \caption{$t = 1926$}
        \label{fig:w8_t4}
    \end{subfigure}
    \hfill
    \begin{subfigure}[b]{0.32\textwidth}
        \centering
        \includegraphics[width=\textwidth]{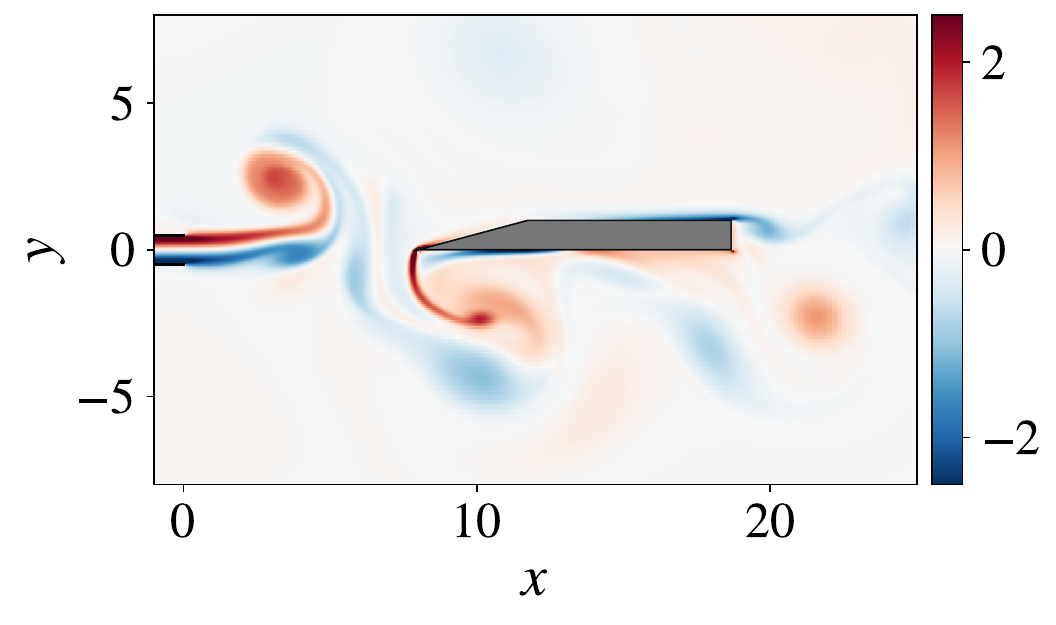}
        \caption{$t = 1940$}
        \label{fig:w8_t5}
    \end{subfigure}
    \hfill
    \begin{subfigure}[b]{0.32\textwidth}
        \centering
        \includegraphics[width=\textwidth]{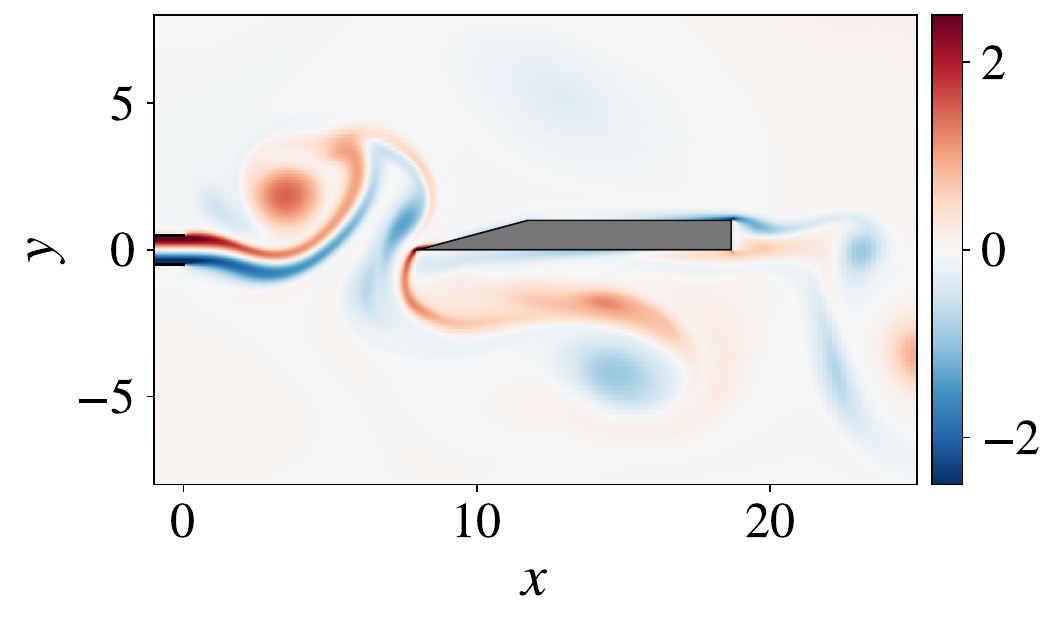}
        \caption{$t = 1953$}
        \label{fig:w8_t6}
    \end{subfigure}

    \caption{Instantaneous vorticity fields at an extended mouth distance of $W = 8$ ($\alpha = 15^{\circ}$, $s = 0$, $Re = 330$), zoomed to the windway exit, jet, and labium region, across six sequential non-dimensional times spanning one collapse-and-recovery cycle of the transverse labium force $F_y(t)$: (a) $t = 1885$, near a local force-amplitude maximum, showing a coherent, well-organized vortex train; (b--c) $t = 1899, 1912$, during a pronounced collapse in oscillation amplitude, where the wake structure becomes visibly weaker and less spatially organized; (d) $t = 1926$, during recovery; (e--f) $t = 1940, 1953$, returning to strong, coherent shedding comparable to (a). This structural cycle accompanies the coexistence of Stage I and the second stage identified in Figure~\ref{fig:FFT_varyW}, and provides flow-field evidence for the amplitude modulation described in the text.}
    \label{fig:vorticity_extended_W}
\end{figure}

The coexistence of the two stages is accompanied by a strong, irregular modulation of the force amplitude, with a collapse and recovery of the oscillation in the interval shown in Figure~\ref{fig:vorticity_extended_W}, which captures the corresponding structural changes in the wake at $W = 8$.

Increasing $W$ thus alters the spectral signature and global dynamics of the system. The visual sequence in Figure~\ref{fig:vorticity_extended_W} is consistent with the spectral picture: the wake alternates between intervals of strong, well-organized shedding and intervals of visibly weaker, less coherent vortex structure, tracking the rise and fall of the force envelope. This behavior is of the type frequently reported in experimental and numerical edge-tone literature~\cite{vaik2014a,paal2007,vaik2011}, in which the transition between discrete operational stages is characterized by the concurrent presence of competing frequencies. The resulting fluid motion is irregular, with substantial amplitude modulation compared to the predominantly periodic $W = 4$ case.

At the maximum studied mouth distance of $W = 12$, the strongest peak, at $St \approx 0.016$, is accompanied by several peaks with about $30$--$40\%$ of its amplitude between $St \approx 0.004$ and $0.056$, and the spectrum is broadened. The force amplitude is highly irregular and the flow field does not repeat periodically over the simulated record.

\begin{figure}[!htbp]
    \centering
    \begin{subfigure}[t]{0.48\textwidth}
        \centering
        \includegraphics[width=\textwidth]{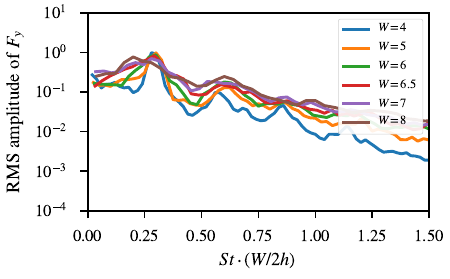}
        \caption{}
        \label{fig:FFT_varyW_rescaled}
    \end{subfigure}
    \hfill
    \begin{subfigure}[t]{0.48\textwidth}
        \centering
        \includegraphics[width=\textwidth]{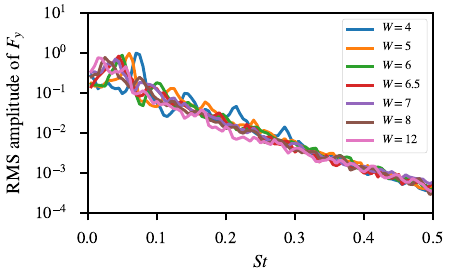}
        \caption{}
        \label{fig:FFT_varyW_raw}
    \end{subfigure}
    \caption{Welch amplitude spectra of the transverse labium force $F_y$ for varying mouth distance $W$ at $Re = 330$, $\alpha = 15^{\circ}$, $s = 0$. (a) Spectra for $W = 4$, $5$, $6$, $6.5$, $7$ and $8$ against $St \cdot (W/2h)$, which uses the mouth distance as the length scale of the feedback delay. The Stage I peaks coincide near $0.29$ up to $W = 7$, and the second-stage peaks appear near $0.2$. (b) Spectra for $W = 4$, $5$, $6$, $6.5$, $7$, $8$ and $12$ against $St$.}
    \label{fig:FFT_varyW}
\end{figure}

Figure~\ref{fig:FFT_varyW}(a) plots the spectra against $St \cdot (W/2h)$, which non-dimensionalizes the frequency with the mouth distance, the distance over which disturbances travel between the windway exit and the labium. A peak whose frequency is set by this distance should appear at a fixed value of this variable, and this is what happens for the Stage I peaks at $W = 4$ to $7$, which coincide near $0.29$. The second-stage peaks appear near $0.2$ and dominate at $W = 8$. Panel (b) shows that, against $St$, the dominant peak moves to lower frequencies as $W$ increases, and that at $W = 12$ the spectrum is broadened. Linear theory instead predicts an exponent of $3/2$: in the present notation, Crighton~\cite{crighton1992} obtained $\omega h/U = (h/W)^{3/2}\,[4\pi(N - 3/8)]^{3/2}$ for stage $N$, with the right dependence on $h$, $U$, $W$ and $N$ but an overall factor that is too large. Because the spectra contain amplitude only, the phase relation between the jet and the labium cannot be assessed here.

The mouth distance $W$ thus controls the fundamental dynamical state of the edge-tone . Rather than merely modulating localized amplitude metrics or inducing a simple linear frequency shift, variations in $W$ change whether the flow operates as a purely periodic, modulated, or irregular system. This finding may be relevant to wind instrument design, although the present model has no resonator, given that the mouth distance relative to the windway height ($W/2h$) remains one of the primary geometric degrees of freedom available to instrument makers. The fact that a tight gap of $W = 4$ (corresponding to a ratio of mouth distance to windway height of $W/2h = 4$, matching the classical recorder proportion) lies in the predominantly periodic single-stage regime, as does $W = 3$, is consistent with the empirical design choices historically adopted by traditional instrument makers~\cite{dequand2003}.

\subsection{Influence of Transverse Jet-to-Labium Offset ($s$)}
\label{subsec:influence_offset}

In practical wind instruments and edge-tone applications, geometric symmetry is frequently broken by design. In flute-like instruments such as the recorder the labium is a single-sided edge, and the windway channel centerline is deliberately shifted relative to the labium tip ($s \neq 0$) to favor reliable asymmetric jet splitting, dictate initial transient receptivity, and suppress unstable regime hopping~\cite{coltman1976,giordano2014}. In organ pipes, the height of the languid, which sets the direction of the jet relative to the upper lip, is adjusted during voicing with regard to the steady tone and the promptness of speech~\cite{fletcher1998}; the offset controls the harmonic content of the jet drive, a symmetric jet generating no even harmonics~\cite{fletcher1998}; and displacing the edge to one side prolongs the lower stage of an edge-tone ~\cite{brown1937}. In the present model the asymmetry is already present at $s = 0$ through the single-sided labium (Section~\ref{sec:methodology}), and the non-dimensional vertical offset $s$ quantifies the additional asymmetry introduced by displacing the jet. Having established that the Reynolds number leaves the oscillation frequency largely unaffected (Section~\ref{subsec:dependence_re}), and that the mouth distance $W$ sets both the oscillation frequency, which scales approximately as $1/W$, and the number of competing stages (Section~\ref{subsec:effect_mouth_distance}), we now ask what breaking the geometric symmetry changes, at the baseline mouth distance $W = 4$. We first examine the frequency content of the force, and then how the flow and the force evolve in time for moderate and for extreme offsets.

\begin{figure}[!htbp]
    \centering
    \includegraphics[width=0.6\textwidth]{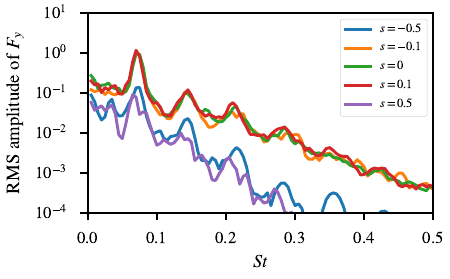}
    \caption{Welch amplitude spectra of the transverse labium force $F_y$ for varying vertical offset $s$, at fixed $W = 4$, $\alpha = 15^{\circ}$, $Re = 330$. The spectra for $s = \pm 0.2$ are omitted for clarity; the corresponding time histories are shown in Figure~\ref{fig:offset_moderate}.}
    \label{fig:FFT_varyS}
\end{figure}

To quantify the spectral modifications at $Re = 330$, Figure~\ref{fig:FFT_varyS} shows the Welch amplitude spectra of the transverse labium force $F_y$ for the different offsets.

All offsets exhibit a primary spectral peak at approximately the same Strouhal number, $St_1 \approx 0.07$, so the offset does not substantially alter the dominant oscillation frequency; the spectra contain amplitude only, so the feedback-loop timing cannot be assessed here. What changes with $s$ is the amplitude. The near-centerline offsets ($s = 0, \pm 0.1$) have practically identical spectra, with prominent energy at the fundamental and its harmonics, consistent with organized vortex splitting at the labium tip, whereas for $|s| = 0.5$ both the peak and the harmonics are much weaker.

\begin{figure}[!htbp]
    \centering
    \begin{subfigure}[b]{0.48\textwidth}
        \centering
        \includegraphics[width=\textwidth]{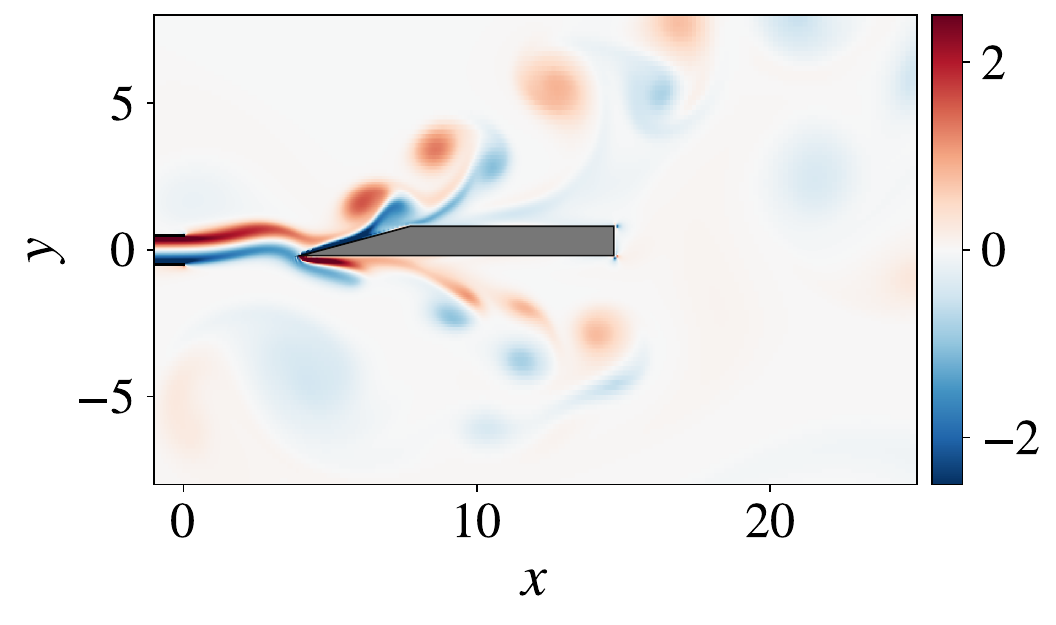}
        \caption{$s = -0.2$}
        \label{fig:s_neg_02}
    \end{subfigure}
    \hfill
    \begin{subfigure}[b]{0.48\textwidth}
        \centering
        \includegraphics[width=\textwidth]{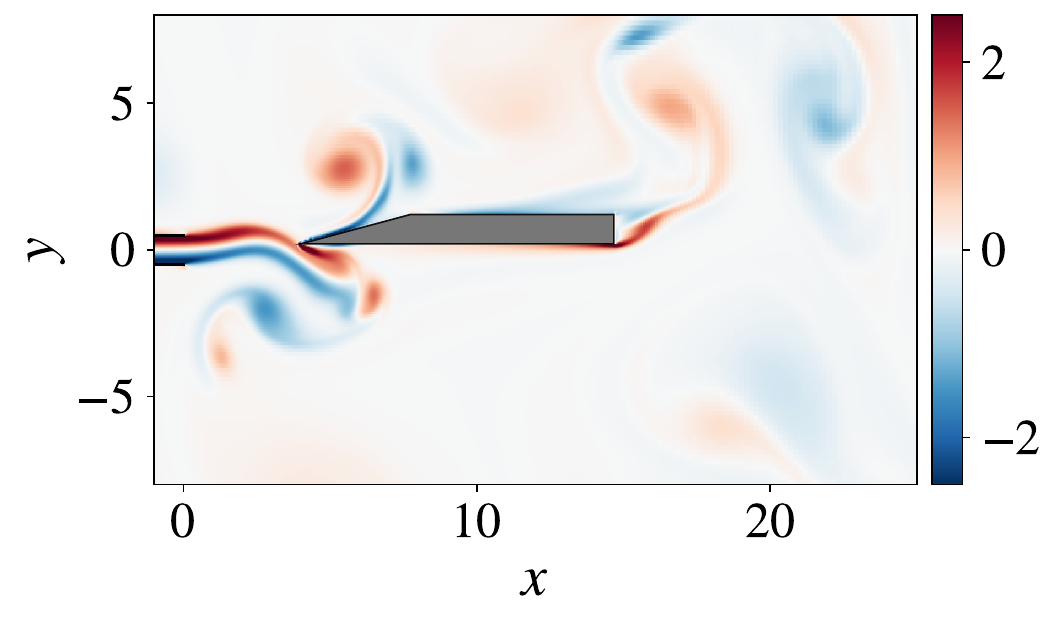}
        \caption{$s = 0.2$}
        \label{fig:s_pos_02}
    \end{subfigure}

    \vspace{0.3cm}

    \begin{subfigure}[b]{\textwidth}
        \centering
        \includegraphics[width=\textwidth]{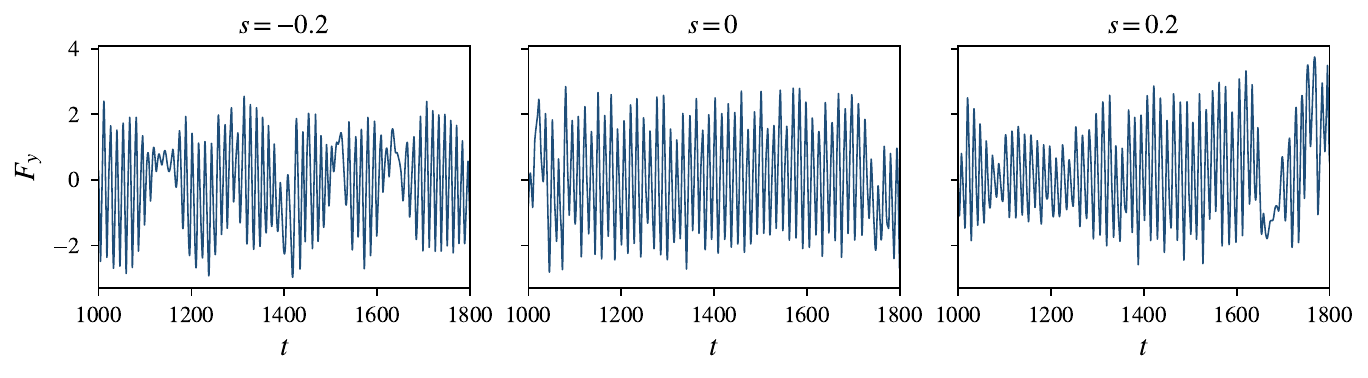}
        \caption{Time history of $F_y$}
        \label{fig:Fy_timehistory_moderate}
    \end{subfigure}

    \caption{Moderate vertical offsets at $Re = 330$ ($W = 4$, $\alpha = 15^{\circ}$). (a--b) Instantaneous vorticity fields at $t = 1800$ for $s = -0.2$ and $s = 0.2$, zoomed to the windway exit, jet, and labium region (red: positive, blue: negative vorticity). (c) Time history of the transverse labium force $F_y$ for $s = -0.2$, $0$ and $0.2$ (left to right), shown for $1000 \le t \le 1800$, the time of the vorticity snapshots.}
    \label{fig:offset_moderate}
\end{figure}

\paragraph{Moderate offsets.}
Figure~\ref{fig:offset_moderate} shows the vorticity field and the force history for $s = \pm 0.2$ and $s = 0$. At these offsets the high-momentum core of the jet remains sufficiently aligned with the labium tip to impinge on the tip rather than being deflected onto the sloped wedge face. The shear layers continue to shed alternating vortical structures, and the shear-layer disturbances are split at the tip, which maintains the hydrodynamic feedback loop across the mouth gap~\cite{powell1961}. The force history in Figure~\ref{fig:offset_moderate}(c) shows a large-amplitude oscillation for all three offsets, with $F_y$ reaching about $\pm 2.5$ to $\pm 3.5$, but the amplitude is not constant. At $s = 0$ it varies only slightly, apart from the episode of one-sided bias near $t \approx 1000$ described in Section~\ref{subsec:overview_regimes} (Figure~\ref{fig:intermittency}). For $s = -0.2$, intervals of strong shedding alternate with weaker intervals, in which the amplitude falls to about $\pm 1$. For $s = 0.2$, the amplitude changes slowly, growing from about $\pm 1.5$ at the start of the interval to about $\pm 3.5$ near $t = 1800$, with a brief weak interval near $t \approx 1650$.

\begin{figure}[!htbp]
    \centering
    \begin{subfigure}[b]{0.48\textwidth}
        \centering
        \includegraphics[width=\textwidth]{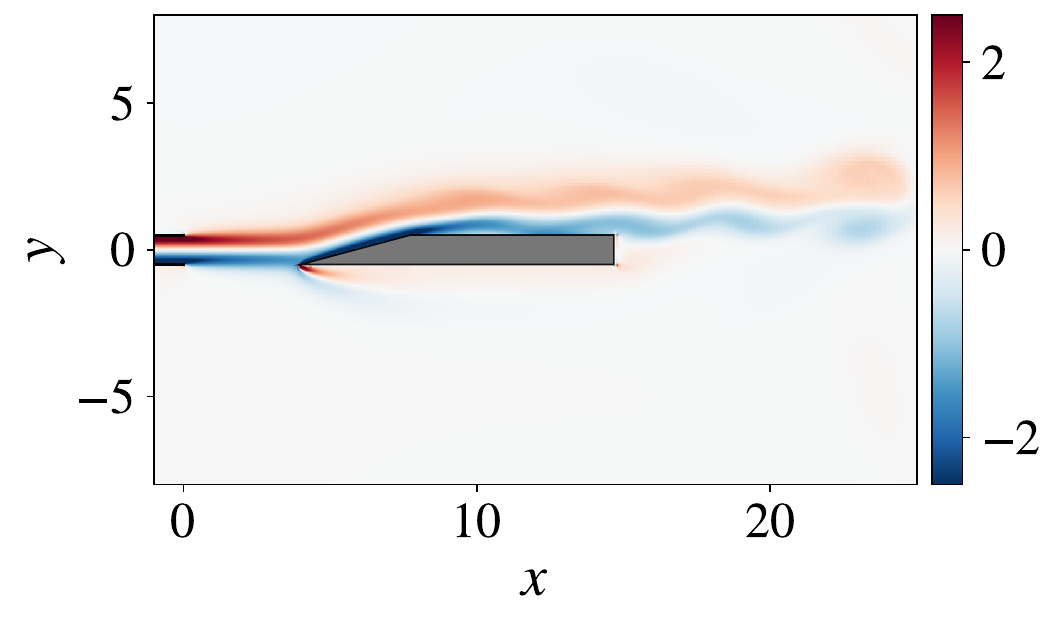}
        \caption{$s = -0.5$}
        \label{fig:s_neg_05}
    \end{subfigure}
    \hfill
    \begin{subfigure}[b]{0.48\textwidth}
        \centering
        \includegraphics[width=\textwidth]{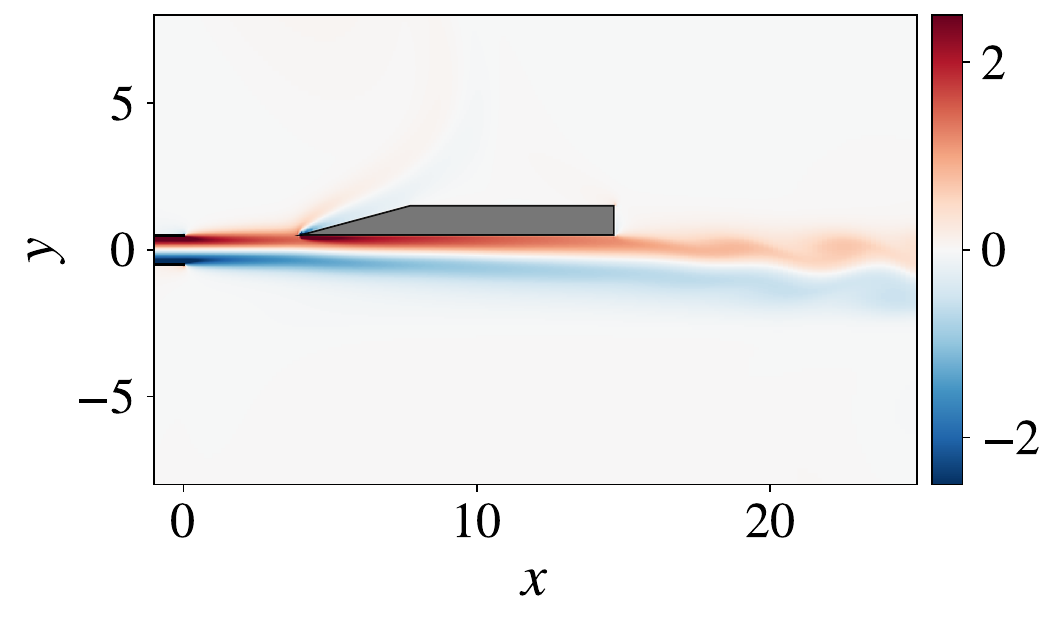}
        \caption{$s = 0.5$}
        \label{fig:s_pos_05}
    \end{subfigure}

    \vspace{0.3cm}

    \begin{subfigure}[b]{0.67\textwidth}
        \centering
        \includegraphics[width=\textwidth]{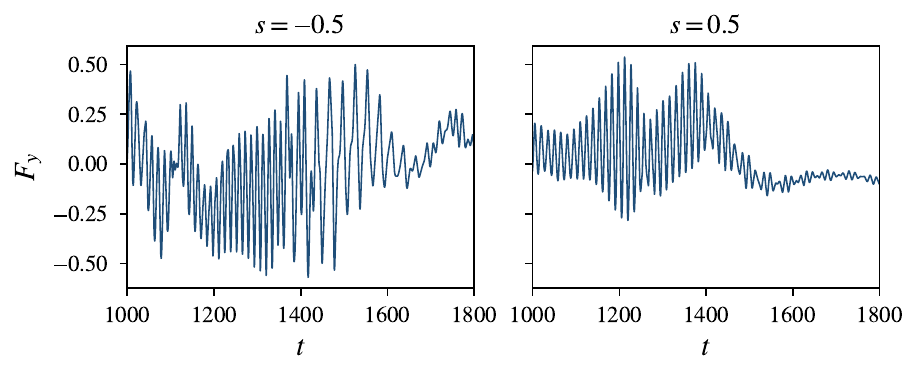}
        \caption{Time history of $F_y$}
        \label{fig:Fy_timehistory_extreme}
    \end{subfigure}

    \caption{Extreme vertical offsets at $Re = 330$ ($W = 4$, $\alpha = 15^{\circ}$). (a--b) Instantaneous vorticity fields at $t = 1800$ for $s = -0.5$ and $s = 0.5$ (red: positive, blue: negative vorticity). (c) Time history of the transverse labium force $F_y$ for $s = -0.5$ and $s = 0.5$ (left to right), shown for $1000 \le t \le 1800$; note the different vertical scale from Figure~\ref{fig:offset_moderate}.}
    \label{fig:offset_extreme}
\end{figure}

\paragraph{Extreme offsets.}
When the offset is increased to $|s| = 0.5$, half a windway height, the highest-velocity core of the jet is displaced away from the sharp labium tip and no longer impinges on the tip, as shown in Figures~\ref{fig:s_neg_05} and \ref{fig:s_pos_05}. The two signs of the offset are not equivalent geometries: for $s = -0.5$ the jet runs along the upper face of the labium, whereas for $s = 0.5$ it passes below the labium. In both cases the shear layers remain smooth at this instant, with no alternating vortex street visible. The force histories in Figure~\ref{fig:offset_extreme}(c) show that the two signs of the offset also behave differently in time, and that in neither case does the oscillation reach the amplitude seen at the moderate offsets: $F_y$ remains within roughly $\pm 0.5$. For $s = 0.5$, a strongly modulated oscillation persists until $t \approx 1450$ and then collapses, leaving a small, almost steady negative force with only weak oscillations over the rest of the interval shown, consistent with the smooth shear layers in Figure~\ref{fig:s_pos_05}. For $s = -0.5$, the oscillation remains irregular throughout the interval shown, and the mean force changes sign over time, consistent with the jet being deflected toward one side of the labium during some intervals and toward the other during others. The loss of impingement at the tip thus weakens the mechanism that sustains the edge-tone , as also seen in the spectra of Figure~\ref{fig:FFT_varyS}, and the resulting behavior depends on the sign of the offset. This strong reduction is specific to the short mouth distances: at $W \ge 5$ the force amplitude at $|s| = 0.5$ remains within about $11\%$ of its value at $s = 0$ (Figure~\ref{fig:regime_maps}).

\subsection{Effect of Labium Angle}

The labium angle $\alpha$ was systematically varied across $15^{\circ}$, $30^{\circ}$, and $60^{\circ}$, ranging from the sharp wedge profiles typical of traditional recorder mouthpieces to the blunter, more obtuse geometries encountered in transverse flutes and organ pipes~\cite{dequand2003}. The baseline case $\alpha = 15^{\circ}$ was examined in detail in Figure~\ref{fig:flow_visualization_baseline}, where six sequential snapshots spanning most of one oscillation period established the reference shedding cycle; the present comparison focuses on the wider angles $\alpha = 30^{\circ}$ and $60^{\circ}$ to assess whether this cycle is preserved as the wedge geometry is blunted.

\begin{figure}[!htbp]
    \centering
    \begin{subfigure}[b]{0.32\textwidth}
        \centering
        \includegraphics[width=\textwidth]{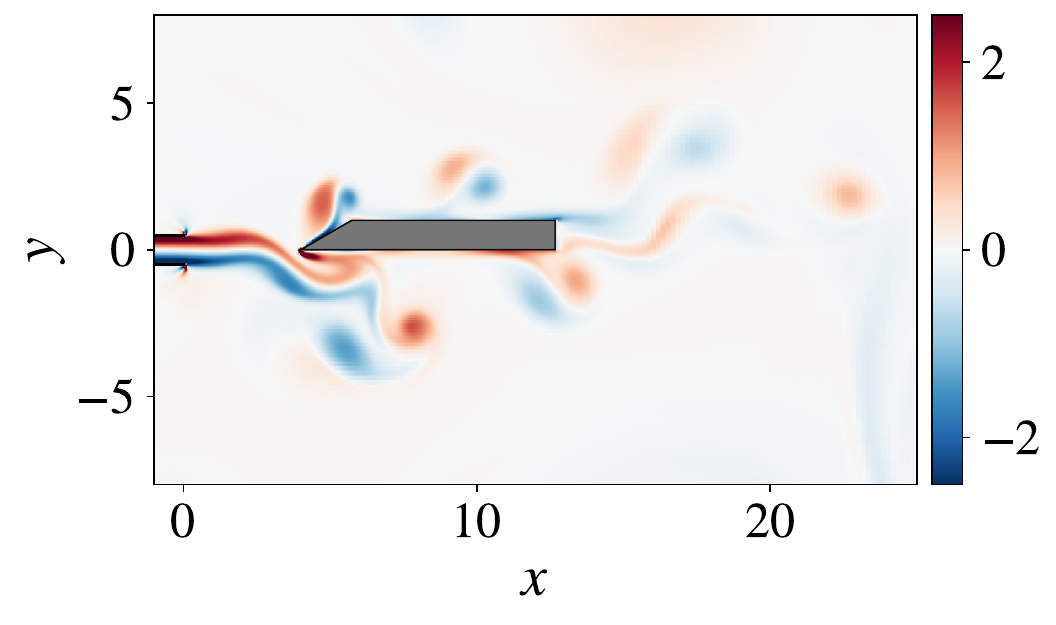}
        \caption{$\alpha = 30^{\circ}$, $t = 1850$}
        \label{fig:vort_a30_t1}
    \end{subfigure}
    \hfill
    \begin{subfigure}[b]{0.32\textwidth}
        \centering
        \includegraphics[width=\textwidth]{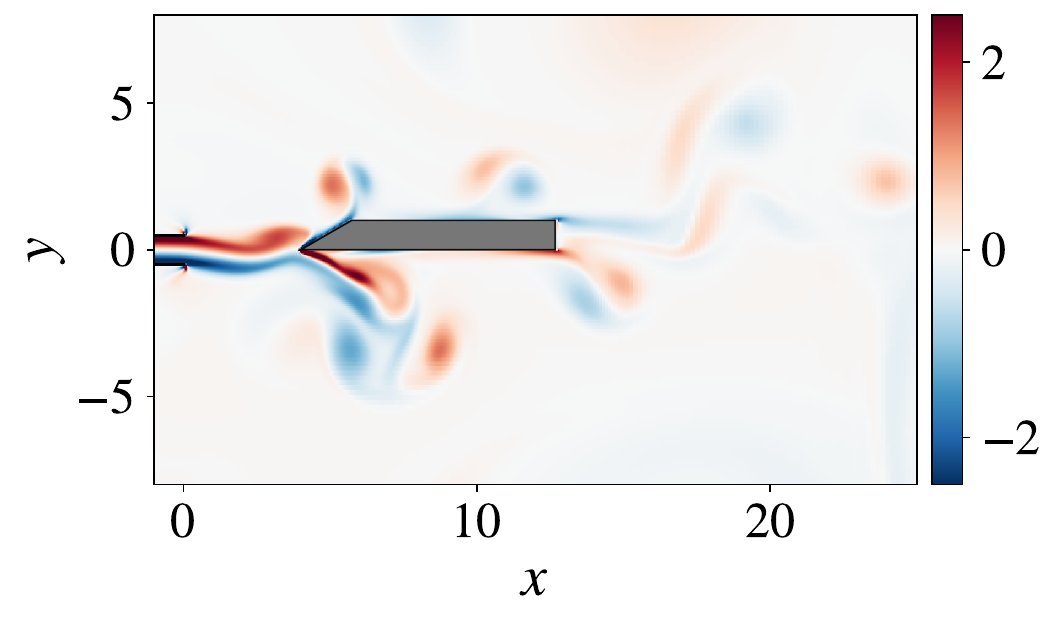}
        \caption{$\alpha = 30^{\circ}$, $t = 1855$}
        \label{fig:vort_a30_t2}
    \end{subfigure}
    \hfill
    \begin{subfigure}[b]{0.32\textwidth}
        \centering
        \includegraphics[width=\textwidth]{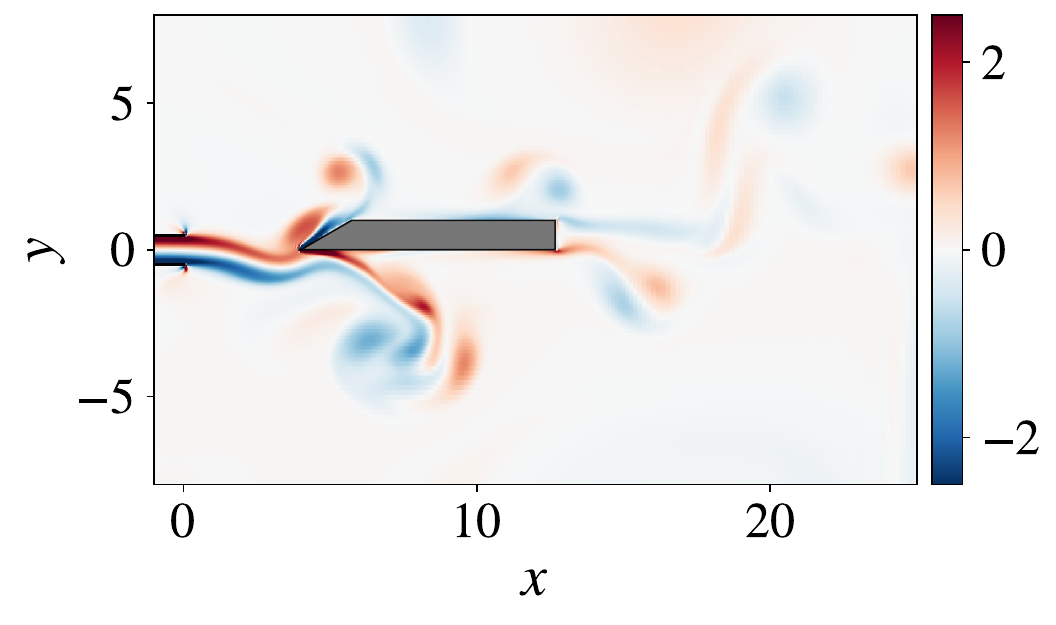}
        \caption{$\alpha = 30^{\circ}$, $t = 1859$}
        \label{fig:vort_a30_t3}
    \end{subfigure}

    \vspace{0.35cm}

    \begin{subfigure}[b]{0.32\textwidth}
        \centering
        \includegraphics[width=\textwidth]{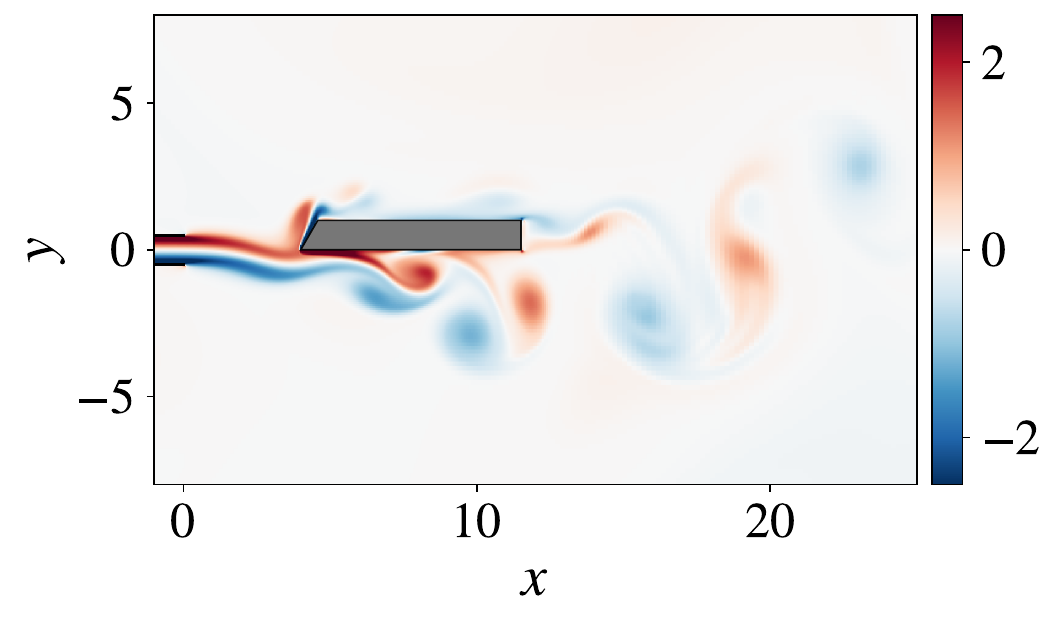}
        \caption{$\alpha = 60^{\circ}$, $t = 1850$}
        \label{fig:vort_a60_t1}
    \end{subfigure}
    \hfill
    \begin{subfigure}[b]{0.32\textwidth}
        \centering
        \includegraphics[width=\textwidth]{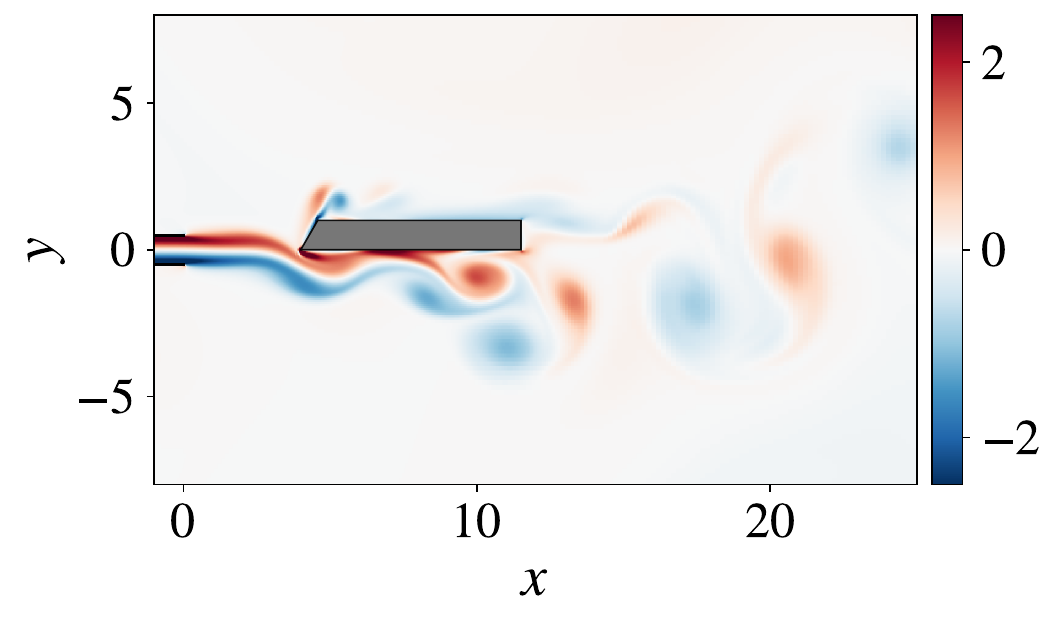}
        \caption{$\alpha = 60^{\circ}$, $t = 1855$}
        \label{fig:vort_a60_t2}
    \end{subfigure}
    \hfill
    \begin{subfigure}[b]{0.32\textwidth}
        \centering
        \includegraphics[width=\textwidth]{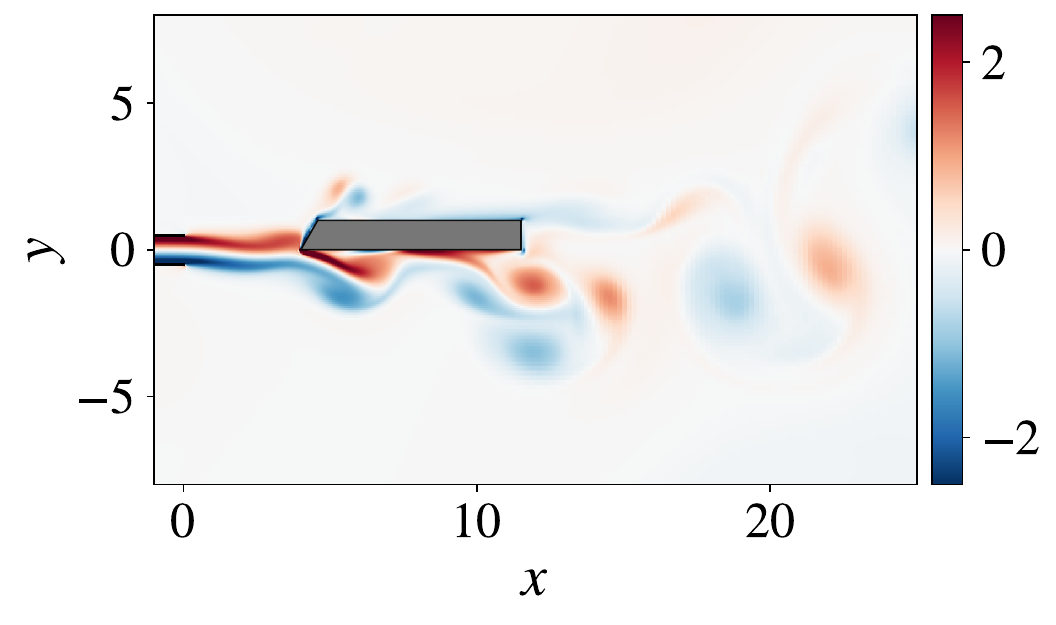}
        \caption{$\alpha = 60^{\circ}$, $t = 1859$}
        \label{fig:vort_a60_t3}
    \end{subfigure}

    \caption{Instantaneous vorticity fields at $Re = 330$ ($W = 4$, $s = 0$), zoomed to the windway exit, jet, and labium region, comparing labium angles $\alpha = 30^{\circ}$ (top row) and $\alpha = 60^{\circ}$ (bottom row) at three sequential non-dimensional times within one oscillation period ($T \approx 1/St_1 \approx 14.3$, with $St_1 \approx 0.07$, the same value obtained for the $\alpha = 15^{\circ}$ baseline in Figure~\ref{fig:flow_visualization_baseline}). The sequence for each angle traces the same phase of shear-layer roll-up, vortex advection, and tip-splitting seen at the baseline angle, indicating that the periodic Stage I edge-tone cycle is preserved as the wedge is blunted.}
    \label{fig:vorticity_varying_angles}
\end{figure}

\begin{figure}[!htbp]
    \centering
    \includegraphics[width=0.6\textwidth]{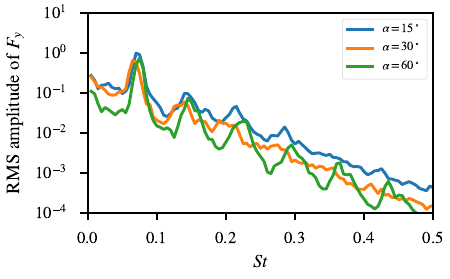}
    \caption{Welch amplitude spectra of the transverse force $F_y$ on the labium wedge across varying inclination angles ($\alpha = 15^{\circ}, 30^{\circ}, 60^{\circ}$) under baseline symmetric conditions ($W = 4$, $Re = 330$, $s = 0$). The close spectral collapse onto a single, dominant primary frequency peak ($St_1 \approx 0.07$) and its harmonics, within the resolution of the simulations, indicates that the wedge angle acts predominantly as an amplitude modulator rather than a frequency-tuning element, and that, within the amplitude-only spectra available here, the fundamental frequency shows no strong sensitivity to wedge sharpness in the range tested.}
    \label{fig:FFT_varyA}
\end{figure}

Figure~\ref{fig:vorticity_varying_angles} shows instantaneous vorticity fields for the two wider wedge angles, each traced across three instants within one oscillation period using the same three-snapshot spacing and zoomed field of view as the $\alpha = 15^{\circ}$ baseline sequence in Figure~\ref{fig:flow_visualization_baseline}. At corresponding phases, the path of the jet is the same for the three angles and remains well centered regardless of wedge inclination, continuing to generate periodic, alternating vortical structures along the upper and lower shear layers at each sampled instant. The cycle of shear-layer roll-up, downstream advection, and tip-splitting at the labium tip observed at $\alpha = 15^{\circ}$ is reproduced essentially unchanged at $\alpha = 30^{\circ}$ and $\alpha = 60^{\circ}$, and the dominant oscillation frequency is unaffected, consistent with the spectral trends shown in Figure~\ref{fig:FFT_varyA}.

At the baseline mouth distance and Reynolds number ($W = 4$, $Re = 330$), the core oscillation regime is preserved when changing only the labium angle: the dominant Strouhal number and the number of competing frequency peaks in the spectrum remain essentially unchanged across the investigated values of $\alpha$. This result indicates that, within the parameter range tested here, the labium tip angle does not govern the convective instability mechanism or the phase-matching timing of the jet--edge interaction loop. Instead, the frequency appears to be set mainly by the mouth distance $W$ and modulated by the symmetry-breaking offset $s$, while $\alpha$ acts as a secondary geometric parameter affecting the flow locally near the wedge tip. Closer inspection of the force magnitudes shows that the sharpest labium angle ($15^{\circ}$) produces a root-mean-square force of $1.35$, about $40\%$ higher than at $30^{\circ}$ ($0.96$) and $50\%$ higher than at $60^{\circ}$ ($0.89$), computed over $1000 \le t \le 2000$ from a single record per case, whereas the difference between $30^{\circ}$ and $60^{\circ}$ is small. This amplitude effect is small compared with the influence of $W$ and $s$.

Physically, this is consistent with classical edge-tone literature~\cite{brown1937b,powell1961}, in which the feedback loop is understood to be completed upstream of the obstacle through the growth of hydrodynamic shear-layer waves, with the absolute geometry of the cutting edge playing a secondary role in setting the frequency provided it remains sufficiently sharp to cause flow separation as the jet impinges~\cite{dequand2003}.

Taken together, the four parameter sweeps show that the mouth distance $W$ has the largest effect on the oscillation regime. It is the only parameter whose variation changes the number of competing oscillation stages: the Reynolds number modulates mainly the saturation amplitude, the labium angle $\alpha$ leaves the dominant frequency and single-stage character unchanged, and the offset $s$, at the baseline mouth distance $W = 4$, strongly reduces the oscillation amplitude at $|s| = 0.5$ without shifting the dominant frequency.

\begin{figure}[!htbp]
    \centering
    \begin{subfigure}[b]{0.49\textwidth}
        \centering
        \includegraphics[width=\textwidth]{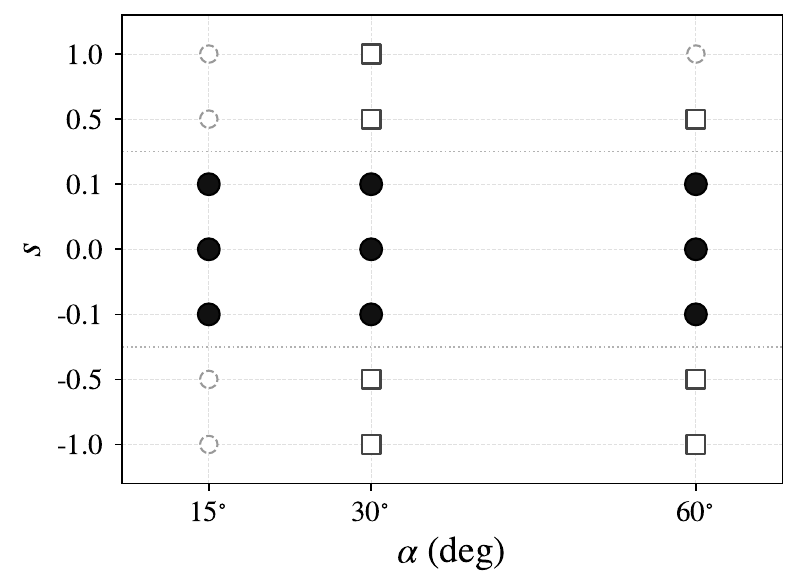}
        \caption{Regime map as a function of $\alpha$ ($W = 4$, $Re = 330$).}
        \label{fig:conclusion_s_vs_a}
    \end{subfigure}
    \hfill
    \begin{subfigure}[b]{0.49\textwidth}
        \centering
        \includegraphics[width=\textwidth]{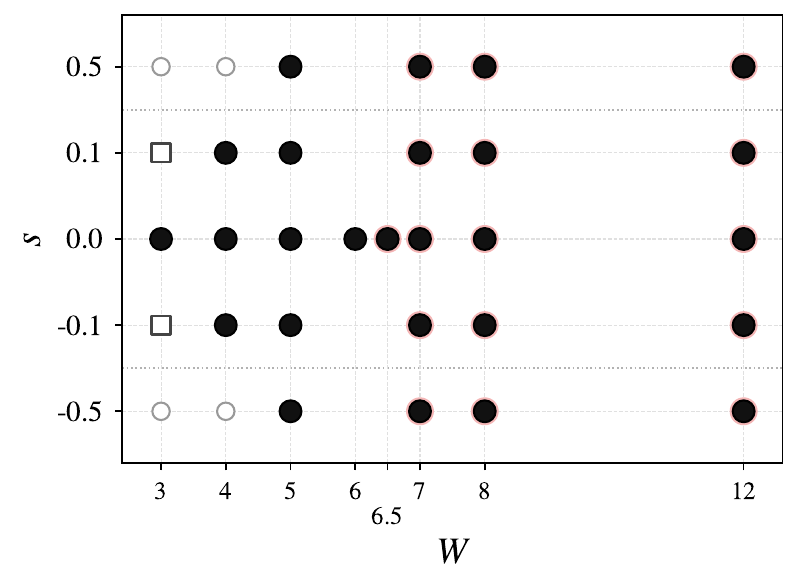}
        \caption{Regime map as a function of $W$ ($\alpha = 15^{\circ}$, $Re = 330$).}
        \label{fig:conclusion_s_vs_w}
    \end{subfigure}

\caption{Hydrodynamic regime maps of the edge-tone mouthpiece at $Re = 330$, showing the transitions between strong, asymmetric, and weak/bypassing oscillation regimes as a function of the vertical offset $s$ and (a) the labium angle $\alpha$ at $W = 4$, (b) the mouth distance $W$ at $\alpha = 15^{\circ}$. \legstrong\ denotes the strong oscillation regime, \legasym\ denotes the asymmetric oscillation regime, and \legbypass\ denotes the weak/bypassing oscillation regime. The regimes are classified from the root-mean-square amplitude of $F_y$ over $t \ge 1000$ relative to the symmetric case ($s = 0$) at the same $\alpha$ (panel a) or the same $W$ (panel b): strong if the ratio exceeds $0.85$, weak/bypassing if it is below $0.35$, and asymmetric in between; the intermediate class is thus defined by amplitude. Dotted horizontal lines at $s = \pm0.35$ separate the strong regime from the weaker regimes; they are drawn between sampled offsets and do not mark a measured threshold. The $s$ axis of both panels and the $\alpha$ axis of panel (a) are categorical (the sampled values are equally spaced), whereas the $W$ axis of panel (b) is numeric. In panel (b), \legsecondmode\ denotes points at $W \ge 6.5$ where the second stage is clearly established, producing strong amplitude modulation; $W = 6$ and $6.5$ were run at $s = 0$ only.}
    \label{fig:regime_maps}
\end{figure}

Figure~\ref{fig:regime_maps} summarizes these regimes in the $(s,\alpha)$ plane at $W = 4$ and in the $(s,W)$ plane at $\alpha = 15^{\circ}$. To check the robustness of the regime boundaries, the offset range was additionally extended to $s = \pm 1$ (not part of the original parametric matrix). At $\alpha = 15^{\circ}$ and $\alpha = 30^{\circ}$, the classified regime at $s = \pm 1$ matches that already observed at $s = \pm 0.5$, so no further regime transition occurs beyond this offset. At $\alpha = 60^{\circ}$, however, the negative offset ($s = -1$) remains in the asymmetric regime, while the positive offset ($s = +1$) is in the bypass regime, unlike the symmetric behavior observed at $s = \pm 0.5$.

\FloatBarrier
\section{Conclusion}
\label{sec:conclusion}

In this work, the hydrodynamic regime characteristics of an edge-tone mouthpiece configuration were systematically mapped across a set of more than 200 direct numerical simulations. By isolating the fluid dynamics from a downstream acoustic resonator, the influence of the Reynolds number, the mouth distance ($W$), the vertical offset ($s$), and the labium angle ($\alpha$) was established. Of the four parameters, the mouth distance $W$ has the largest effect on the oscillation regime: it is the only parameter whose variation changes the number of competing oscillation stages. The resulting regime maps (Figure~\ref{fig:regime_maps}) separate predominantly periodic edge-tone s, intermittent and asymmetric oscillations, and weak, jet-bypass regimes.

The key physical insights governing the hydrodynamic feedback loop are synthesized as follows:

\begin{itemize}
\item \textbf{Role of Reynolds Number ($Re$):} For $Re = 330$, $500$ and $700$ the fundamental Strouhal number remains nearly constant and the spectral structure is preserved, so the flow stays in Stage I. At $Re = 200$ the oscillation is sustained only for near-centerline offsets.

\item \textbf{Dominance of Mouth Distance ($W$) and Stage Bifurcation:} The mouth distance $W$ is the main parameter of the feedback loop, since it sets the convective time delay of shear-layer disturbances between the windway exit and the labium. For $W = 3$ and $4$ the system locks into a single-stage, predominantly periodic oscillation (Stage I), and the frequency of the Stage I peak follows $St \cdot (W/2h) \approx 0.29$, a scaling that it keeps up to $W = 7$. A second stage, with peaks near $St \cdot (W/2h) \approx 0.2$, is only weakly present at $W = 5$ and $6$ (about $20\%$ of the amplitude of the Stage I peak), clearly established from $W = 6.5$, where the Stage I peak still has the highest amplitude, and dominant for $W \ge 8$, where it produces strong amplitude modulation. The onset of the second stage is therefore gradual. At the largest mouth distance studied, $W = 12$, the spectra are broadened and the force is highly irregular.

\item \textbf{Role of Spatial Offset ($s$):} At the baseline mouth distance $W = 4$, the fundamental Strouhal number remains $St_1 \approx 0.07$ for the near-centerline offsets, and the spectra retain prominent energy at the fundamental and its harmonics. For the small offsets $s = \pm 0.1$ the spectra are practically indistinguishable from the symmetric case. The time histories for $s = 0, \pm 0.2$ show a strong oscillation that is less variable at $s = 0$ and more variable in amplitude at $s = \pm 0.2$. At the largest offset, $|s| = 0.5$, the high-momentum core of the jet no longer impinges on the labium tip and the oscillation amplitude is strongly reduced, almost extinguished for $s = 0.5$, with a different behavior for each sign of the offset. This reduction is specific to the short mouth distances.

\item \textbf{Role of Labium Angle ($\alpha$):} At the baseline mouth distance and Reynolds number, the dominant Strouhal number and the number of competing frequency peaks remain essentially unchanged across $\alpha = 15^{\circ}$, $30^{\circ}$ and $60^{\circ}$, so the labium angle does not govern the convective instability mechanism. It acts as a secondary geometric parameter that affects the flow locally near the labium tip, with a higher force amplitude for the sharpest labium.
\end{itemize}

In summary, the regime maps indicate that, at $W = 4$, near-centerline alignment of the jet with the labium sustains the strongest oscillation, while the mouth distance $W$ dictates whether the system maintains a locked single-tone oscillation or undergoes multi-stage interactions. These transition boundaries describe the hydrodynamic mechanism without a resonator; whether they carry over to pitch stability and the suppression of secondary tones in complete instruments remains to be tested with acoustic coupling.

Taken together, these results support $W = 4$, $\alpha = 15^{\circ}$, $s = 0$, and $Re = 330$ as a natural baseline for future work: it lies within the predominantly periodic Stage I regime, avoids the strong amplitude reduction seen at large offsets, and matches the geometry of a traditional recorder mouthpiece. This configuration offers an efficient starting point for subsequent investigations of the aeroacoustic source mechanism, including the unsteady labium force, the source of dipole sound in a compressible setting, and its eventual coupling to a resonating instrument cavity.

\begin{acknowledgments}
This work is part of the R\&D project PID2022-141243NA-I00, funded by MICIU/AEI/10.13039/\allowbreak501100011033 and by ERDF/EU. A.~T.~is funded through the University of C\'adiz's plan propio. P.~R.\ acknowledges support from DGAPA-UNAM through grant PAPIIT IN114227.

\textit{AI disclosure:} Claude (Anthropic) was used to assist in the writing of the text.
\end{acknowledgments}

\section*{Data availability}
The numerical code used in this study (IBMOS) is open source~\cite{fosasdepando2020ibmos}.

\FloatBarrier
\appendix

\section{Verification of numerical parameters}
\label{app:convergence}

To check that the reported results are independent of domain truncation, of the treatment of the outflow boundary, and of the spatial mesh resolution, convergence studies of the downstream domain length $L_{xd}$, of the sponge layer, and of the mesh resolution were performed at the baseline configuration ($Re = 330$, $W = 4$, $\alpha = 15^\circ$, $s = 0$). Convergence was assessed by comparing the Welch amplitude spectra of the transverse labium force $F_y$, extracted from the saturated oscillation regime.

\subsection{Domain Size and Sponge Layer}
\label{app:domain}

\begin{figure}[!htbp]
\centering
\begin{subfigure}[b]{0.49\textwidth}
\centering
\includegraphics[width=\textwidth]{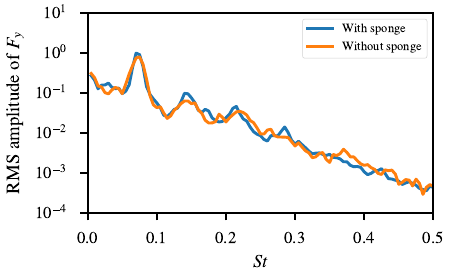}
\caption{With and without the sponge layer.}
\label{fig:sponge_effect}
\end{subfigure}
\hfill
\begin{subfigure}[b]{0.49\textwidth}
\centering
\includegraphics[width=\textwidth]{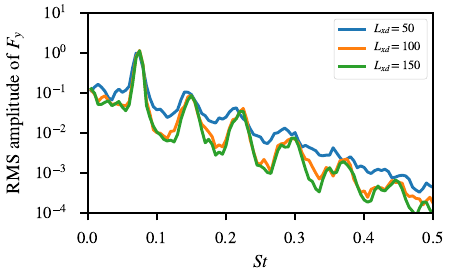}
\caption{Varying downstream domain length $L_{xd}$.}
\label{fig:box_size}
\end{subfigure}
\caption{Welch amplitude spectra of the transverse labium force $F_y$ at the baseline condition ($Re=330$, $W=4$, $\alpha=15^\circ$, $s=0$), assessing sensitivity to outlet-boundary treatment. (a) With and without the sponge layer active. (b) Across $L_{xd} = 50, 100, 150$.}
\label{fig:convergence_domain}
\end{figure}

The influence of the downstream domain length $L_{xd}$ and of the sponge layer was assessed at the baseline flow condition. Figure~\ref{fig:sponge_effect} compares the Welch amplitude spectrum of the transverse labium force $F_y$ with and without the sponge layer active. The dominant spectral peaks, including the fundamental Strouhal number $St_1 \approx 0.07$ and its harmonics, are unaffected by the presence of the sponge layer; the two spectra differ only in the high-frequency spectral content ($St \gtrsim 0.25$), which is suppressed when the sponge layer is active, consistent with the sponge layer suppressing spurious boundary-reflected energy without altering the physical oscillation.

Figure~\ref{fig:box_size} shows the corresponding spectra for three downstream domain lengths, $L_{xd} = 50$, $100$ and $150$. Since the transverse half-height $L_y = 0.5(L_{xd} + L_{xu})$ varies together with $L_{xd}$, this test checks the combined effect of the downstream length and of the transverse size of the domain; the two were not varied separately. The dominant peak and harmonic structure are essentially unchanged across the three domain sizes, so the fundamental oscillation frequency is independent of the domain size, and $L_{xd} = 50$, used throughout this study, is sufficient for the quantities reported here.

\subsection{Mesh Resolution}
\label{app:mesh}

\begin{figure}[!htbp]
\centering
\includegraphics[width=0.6\textwidth]{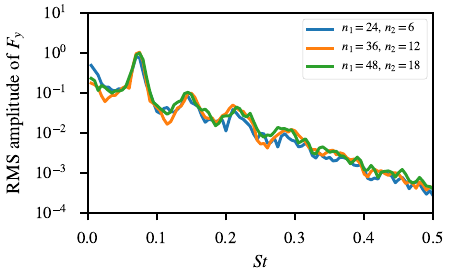}
\caption{Welch amplitude spectra of the transverse labium force $F_y$ at the baseline condition ($Re=330$, $W=4$, $\alpha=15^\circ$, $s=0$) for three near-field/far-field mesh resolutions, $(n_1,n_2) = (24,6)$, $(36,12)$, and $(48,18)$.}
\label{fig:mesh_spectra}
\end{figure}

To check that the results do not depend on the spatial resolution, the baseline case was repeated on three meshes with increasing near-field and far-field resolution, $(n_1, n_2) = (24, 6)$, $(36, 12)$, and $(48, 18)$, where $(24, 6)$ is the resolution used throughout this study (Section~\ref{subsec:resolution}). Both $n_1$ and $n_2$ were refined together. Figure~\ref{fig:mesh_spectra} compares the Welch amplitude spectra of $F_y$ for the three meshes. The three spectra are nearly indistinguishable: the dominant peak remains at $St_1 \approx 0.07$, its harmonics appear at the same frequencies, and the overall spectral shape is preserved. Refining the mesh by a factor of two in each direction therefore does not change the oscillation frequency or the spectral content of the force at the baseline configuration. The mesh study was not repeated for $W \ge 7$ or for $Re = 700$, which uses $n_2 = 12$, so the amplitude differences between $Re = 500$ and $Re = 700$ (Section~\ref{subsec:dependence_re}) are not interpreted.

\FloatBarrier
\bibliographystyle{apsrev4-2}
\bibliography{ref_prfluids}
\end{document}